\documentstyle[12pt]{article}
\newcommand{\beq}{\begin{equation}}
\newcommand{\eeq}{\end{equation}}

\def\quaft{{\textstyle {{{1}\over{4\pi\alpha'}}} }}
\def\half{{\textstyle{1\over2}}}

\def\p1half{{\textstyle{{{p+1}\over{2}}}}}

\def\23phalf{{\textstyle{{{23-p}\over{2}}}}}

\begin{document}
\begin{titlepage}

\bigskip
\hskip 5.2in{\vbox{\baselineskip12pt}}

\bigskip\bigskip\bigskip\bigskip
\centerline{\large\bf The Normalization of Perturbative String
Amplitudes:} \centerline{\large\bf Weyl Covariance and Zeta
Function Regularization}
\bigskip\bigskip
\bigskip\bigskip
\centerline{\bf Shyamoli Chaudhuri\footnote{Current Address: 1312
Oak Dr, Blacksburg, VA 24060. Email: shyamolic@yahoo.com}}
\centerline{214 North Allegheny Street} \centerline{Bellefonte, PA
16823}
\medskip
\date{\today}

\bigskip\bigskip
\begin{abstract}
This is a self-contained pedagogical review of Polchinski's 1986
analysis from first principles of the Polyakov path integral based
on Hawking's zeta function regularization technique for
scale-invariant computations in two-dimensional quantum gravity,
an approach that can be adapted to any of the perturbative string
theories. In particular, we point out the physical significance of
preserving both Weyl and global diffeomorphism invariance while
taking the low energy field theory limit of scattering amplitudes
in an open and closed string theory, giving a brief discussion of
some physics applications. We review the path integral computation
of the pointlike off-shell closed bosonic string propagator due to
Cohen, Moore, Nelson, and Polchinski. The extension of their
methodology to the case of the macroscopic loop propagator in an
embedding flat spacetime geometry has been given by Chaudhuri,
Chen, and Novak. We examine the macroscopic loop amplitude from
the perspective of both the target spacetime massive type II
supergravity theory, and the boundary state formalism of the
worldsheet conformal field theory, clarifying the precise evidence
it provides for a Dirichlet (-2)brane, an identification made by
Chaudhuri. The appendices contain a comprehensive presentation of
the covariant path integral technique for the one loop amplitudes
of the supersymmetric, and unoriented, open and closed type I
string theory, and in an external two-form background field of
generic strength.
\end{abstract}
\noindent

\end{titlepage}

\section{Introduction}

\vskip 0.1in The bread-and-butter tool of high-energy theorists,
whether in particle physics or cosmology, is effective quantum
field theory: a spacetime Lagrangian holding at some given mass
scale $m_e$ with, in general, non-renormalizable corrections
accompanied by inverse powers of $m_s$. Many computations in
string theory also begin from the perspective of an effective
field theory, namely, a ten-dimensional supergravity-Yang-Mills
theory defined at the string scale, $m_s$. An important
distinction from generic effective field theories is that a
detailed worldsheet prescription exists, at least in principle,
for the computation of the nonrenormalizable terms in the string
theory spacetime Lagrangian, and that it holds {\em to all orders
in $m_s$}. Note that, roughly speaking, it is conventional to
identify $m_s$ $=$ $\alpha^{\prime - 1/2}$, where
$\alpha^{\prime}$ is the inverse closed string tension, with the
ten-dimensional Planck scale, $M_P$, as befitting a theory of
quantum gravity. But it should be kept in mind that the effective
four-dimensional string scale can be considerably lower in {\em
specific} vacua of the theory, due to a dependence on either
geometric moduli, or on the background fields and fluxes
characterizing the spacetime geometry. This feature has become
popular in recent phenomenological model building.

\vskip 0.1in In this paper, we wish to explore a key feature that
distinguishes perturbative string theory from generic effective
field theories. This distinction was already highlighted in
Polchinski's 1986 observation \cite{poltorus} that the vacuum
energy in string theory is {\em unambiguously normalized},
inclusive of numerical factors, in terms of just two parameters:
the string scale, $m_s$, and the dimensionless string coupling, or
dilaton vev, $g$$=$$e^{\phi_0}$. Since this is quite unlike the
expectation in generic effective quantum field theories, the
significance of this observation should be clearly appreciated
prior to serious investigation of low energy cosmo-particle
physics in String/M Theory. We should also note that this result
has crucial consequences for the thermodynamics of the canonical
ensemble of perturbative strings \cite{poltorus,holo}. The vacuum
energy of perturbative strings in a flat spacetime geometry is the
most fundamental dimensionful quantity we would think to compute
in string theory. But it should be emphasized that even when the
vacuum energy receives additional, tree-level corrections from
possible Dbranes, or from a background two-form field strength, as
in many braneworld models of the type I and type II string
theories, it is nonetheless true that such mass contributions are
unambiguously computible in terms of the single mass scale $m_s$,
plus a specified number of dimensionless parameters that include
the string coupling, or dilaton vev. This second remarkable
observation follows from Polchinski's 1995 worldsheet computation
of the Dbrane tension in type I string theory \cite{dbrane}. In
the next three sections, and in appendices A thru C, we will
provide a pedagogical review of some relevant technical details
and background that are helpful in understanding the derivation of
these two key results.

\vskip 0.1in In section 5, we review an additional far-reaching
insight, first noticed by us during the course of the work
reported on in Ref.\ \cite{ncom}. Our observation follows by
adapting the arguments that led to the worldsheet computation of
the Dbrane tension \cite{dbrane} to the {\em opposite} limit of
the annulus graph, dominated by the lowest-lying open string modes
in the worldvolume gauge theory. We will find that the remarkable
normalizability property of the open string scattering amplitude
survives in the low energy effective field theory limit: it is
possible to determine all of the couplings in the worldvolume
effective Lagrangian, inclusive of numerical factors, in terms of
the single mass scale, $m_s$, plus a finite number of
dimensionless scalar field expectation values, or moduli. Our
observation exploits an insufficiently exploited property of the
type I and type I$^{\prime}$ open string theories, first noted by
us in \cite{ncom}. Consider the one-loop scattering amplitudes. In
order to extract the contribution from the lowest-lying open
string modes dominating the short cylinder limit of some given
scattering amplitude, there is no need to set the modulus of the
annulus to some ad-hoc value, thereby explicitly breaking
reparametrization invariance. Instead, we can first expose the
asymptotics of the integrand by Taylor expansion, followed by
explicit evaluation of the modular integral. The physical
significance of having thus preserved all of the manifest
symmetries of the worldsheet formalism, namely, exact Weyl and
diffeomorphism invariance, is that we obtain an {\em unambiguously
normalized} expression for each of the couplings in the
worldvolume effective field theory (EFT). Each is derived directly
from an unambiguously normalized perturbative string scattering
amplitude \cite{ncom}. This observation could have significant
consequences for the derivation of perturbative gauge theory
results from perturbative string scattering amplitude
computations.

\vskip 0.1in Finally, in section 6, and in appendix D, we review
the path integral computation of the pointlike off-shell closed
bosonic string propagator due to Cohen, Moore, Nelson, and
Polchinski \cite{cmnp}. The extension of their methodology to the
case of macroscopic boundary loops in an embedding flat spacetime
geometry has been given by Chaudhuri, Chen, and Novak \cite{wils}.
We explain this modification of the familiar one-loop string
amplitude from the perspective of the boundary state formalism of
the worldsheet conformal field theory, pointing out that it gives
concrete worldsheet evidence for a Dirichlet (-2)brane boundary
state in string theory. The consequences for the Dpbrane spectrum
of String/M theory will be explored elsewhere. We should note that
in the case of more complicated loop geometries, including the
interesting possibility of {\em corners} \cite{wils}, there exist
a number of potential physics applications for macroscopic loop
amplitudes both in cosmology, and in condensed matter physics.
These relate to cosmic string production from the vacuum, cosmic
string scattering, as well as the study of radiation from cusps
and/or corners on a cosmic string. It would be remarkable to
derive such phenomena directly from a fundamental, and fully
renormalizable, quantum theory.

\vskip 0.1in This paper contains a topical review of pedagogical
material which we feel is, regretfully, unavailable in any of the
standard string theory textbooks and reviews. It should be useful
to both students, and to practitioners of effective field theory,
who wish to better understand the full import of the worldsheet
approach to string theory. It should also be helpful to string
theorists unfamiliar with the strengths of the covariant path
integral approach. We have tried to provide an equivalent
discussion in the operator formalism whenever this is absent in
the current literature, as in our discussion in section 6 of the
D(-2)brane boundary state. More generally, a description of the
operator formulation for the type IB and type I$^{\prime}$ string
theories, and of boundary conformal field theory techniques, can
be easily found in the standard sources in the literature.

\section{The Normalization of String Amplitudes}

\vskip 0.1in Let us begin by reviewing the computation of the
vacuum functional in quantum field theory. In order to enable a
clearer comparison, we will work in the Feynman path integral
formulation for a perturbative quantum field theory. The vacuum
functional of quantum field theory is given by the sum over
classical field configurations, with the natural requirement that
{\em the measure of the path integral preserve the continuous
symmetries of the classical action}. In the case of a non-abelian
gauge theory coupled to a massless scalar field, for example, we
define a gauge invariant measure, factor out the volume of the
gauge group by choosing a slice in field space that intersects
each gauge orbit exactly once, and eliminate the overall infinity
introduced by the choice of a redundant gauge-invariant measure by
dividing by the volume of the gauge group:
\begin{equation}
Z[A,\Phi] = {{1}\over{{\rm Vol[gauge]} }}
    \int [d A] [d \Phi] e^{-S_{YM} [A] -S[\Phi,A] } = \int_{\rm orbit} [d{\hat{A}}]
[d\Phi] e^{-S[{\hat{A}}] + S_{\rm FP} - S[\Phi,{\hat{A}}] }
\quad ,
\label{eq:gauge}
\end{equation}
Although this gives an elegant starting point for calculations in perturbative non-abelian
gauge theory, the vacuum functional of the gauge theory is {\em not} normalizable because
of an overall infinity
introduced by the ultraviolet divergent vacuum energy density. Despite having removed
the redundancy due to gauge invariance by implementing the Faddeev-Popov procedure, the
right-hand-side of Eq.\ (\ref{eq:gauge}) only becomes well-defined
{\em when we introduce an appropriate ultraviolet regulator for the
gauge theory}. This property is exactly the same as in any other quantum field theory,
whether renormalizable, or not. Thus, in order to obtain physically meaningful results which hold
independent of the choice of ultraviolet regulator, we restrict ourselves in quantum field
theory to computing {\em normalized} correlation functions defined as follows:
\begin{equation}
<O_1 [A,\Phi] \cdots O_n [A,\Phi] > = {{1}\over{Z[A,\Phi]}} {{1}\over{{\rm Vol[gauge]} }}
    \int [d A] [d\Phi] O_1 [A,\Phi] \cdots O_n [A,\Phi] e^{-S_{YM} [A] - S[\Phi,A]}
\quad .
\label{eq:gaugecor}
\end{equation}
The ambiguity introduced by the introduction of an ultraviolet
regulator in both the numerator and denominator has been {\em
cancelled} by taking the ratio. A suitable choice of regulator in
this example would be dimensional regularization, which preserves
the non-abelian gauge invariance. We assume the reader is
well-acquainted with the remarkably successful computational
scheme for perturbative Yang-Mills theory that follows from this
prescription. However, as is well-known, the cosmological constant
problem has not been addressed in this field theoretic analysis:
the vacuum energy density is regulator-dependent, formally
infinite, or at least of the same order as the overall mass scale
of the quantum field theory.

\vskip 0.1in
The path integral for two-dimensional quantum gravity coupled to $d$ massless scalar fields,
namely, the worldsheet action for $d$-dimensional bosonic string theory, can be analysed
exactly along these same lines \cite{polyakov,poltorus}. Two-dimensional quantum gravity is
almost pure gauge, as one might expect from the fact that the Einstein action in
two-dimensions is a topological invariant equal to the Euler number of the
two-dimensional manifold:
\begin{equation}
S[g] = \int_M d^2 \xi {\sqrt{-g}} R + 2 \int_{\partial M}
ds k = 4\pi \chi_M  = 4 \pi (2 -2h - b -c) \quad ,
\label{eq:einst}
\end{equation}
where $h$, $b$, and $c$, are, respectively, the number of handles, boundaries, and
cross-caps on the two-dimensional Riemannian manifold. $R$ and $k$ are, respectively,
the Ricci scalar curvature, and the geodesic curvature on the boundary.
However, as was shown by Polyakov in 1983 \cite{polyakov} by a careful analysis of the
Faddeev-Popov functional determinant, in any sub-critical spacetime dimension, $d$ $<$ $26$,
the Weyl mode, $\phi$, in the two-dimensional metric, $g_{ab}$ = $e^{-\phi} {\hat{g}}_{ab}$,
acquires non-trivial dynamics given by the Liouville field theory:
\begin{equation}
Z[g,X] = {{1}\over{{\rm Vol[Diff_0 ]} }}
    \int [d g] [d X] e^{-\lambda S [g] -S[X,g] } = \int_{\rm orbit} [d \phi]
[d X ] e^{-\lambda S[{\hat{g}}] + (d-26) S_{\rm L}[\phi] - S[X, {\hat{g}}] }
\quad ,
\label{eq:path}
\end{equation}
where $\lambda$ is an arbitrary constant that fixes the loop
expansion parameter for two-dimensional quantum gravity:
$e^{-\lambda \chi_M}$. In string theory, $\lambda$ will be
determined by the dimensionless closed string coupling,
$\lambda$$=$$e^{-\Phi_0}$. The notation Diff$_0$ denotes the group
of connected diffeomorphisms of the worldsheet metric. Notice that
the coefficient of the Liouville action vanishes in the critical
spacetime dimension, $d$$=$$26$. Thus, in critical string theory,
$\phi$ drops out of the classical action, and we must divide by
the volume of the Weyl group in order to eliminate the redundancy
in the measure due to Weyl invariance:\footnote{It should be
emphasized that the path integral for two-dimensional
Weyl-invariant quantum gravity computes the sum over {\em
connected} vacuum graphs in string theory. We denote this quantity
by the usual symbol $W$ \cite{poltorus}.}
\begin{eqnarray}
W_{\rm string} [g,X] =&& {{1}\over{{\rm Vol[Diff_0 \times Weyl ]} }}
    \int [d g] [d X] e^{-\lambda S [g] -S[X,g] } \cr
 =&& \int [d\tau]_{\chi_M} e^{-\lambda S[{\hat{g}}]} \int [d X ]
   e^{ - S[X, {\hat{g}}] }
 \equiv \int [d\tau]_{\chi_M}
                          e^{ -S[{\hat{g}}] } ({\rm det}^{\prime} \Delta)^{-(d-2)/2}
\quad .
\label{eq:pathw}
\end{eqnarray}
The path integral simplifies to an ordinary finite-dimensional integral over the
moduli of the worldsheet metric, parameters that describe the shape and topology of
the Riemannian manifold. The first-principles derivation of this result is reviewed
in the
appendix following \cite{poltorus,dhoker,mine}. $\Delta$ denotes the Laplacian acting
on two-dimensional scalars on a Riemann surface with Euler number $\chi_M$ and
moduli, $\tau$. The measure for moduli, $[d\tau]$, can be unambiguously determined
by the requirement
of gauge invariance, as was shown in \cite{poltorus,dhoker,mine}. What remains is
to obtain an expression for the functional determinant on the right-hand-side of the
equation, in as explicit a form as is feasible. This requires specification of an
ultraviolet regulator for the two-dimensional quantum field theory.

\vskip 0.1in The beauty of Weyl-invariant two-dimensional quantum
gravity is that there is {\em only one} choice of ultraviolet
regulator that preserves all of the gauge invariances of the
theory, namely, both connected diffeomorphisms and Weyl
transformations of the metric. That regularization scheme is
zeta-function regularization \cite{zeta}, as was pointed out by
Polchinski in \cite{poltorus}, and it gives, therefore, an
unambiguously normalized expression for both the vacuum functional
of string theory, as well as, in principle, all of the loop
correlation functions at arbitrary order in the perturbation
expansion \cite{ncom}. In practice, other than on worldsurfaces of
vanishing Euler number which contribute to the one-loop amplitudes
of string theory, the eigenspectrum of the scalar Laplacian is
insufficiently well-known to enable explicit calculation of the
functional determinant \cite{dhoker,mine}. But the fact remains
that, unlike what happens in quantum field theory, there is {\em
no ambiguity in the normalization of generic loop correlation
functions introduced by an arbitrariness in the choice of
regulator}.

\vskip 0.1in
Let us recall the basic idea underlying the zeta function regularization of the
functional determinant of a differential operator in a Euclidean quantum field theory
\cite{zeta}. Hawking begins by noting that the functional determinant of a
second-order differential operator $\Delta$ on a generic
$k$-dimensional manifold $M$ with a discrete eigenvalue spectrum, $\{\lambda_n\}$, and
normalized eigenfunctions, $\{\Psi_n\}$, can be interpreted as the generalization of an
ordinary Riemann zeta function. This generalized zeta function
is, formally, given by the sum over the eigenvalues of $\Delta$:
\begin{equation}
\Delta \Psi_n = \lambda_n \Psi_n , \quad \int \Psi_n \Psi_m {\sqrt{-g}} ~ d^k \xi
= C_n \delta_{mn} , \quad \Phi = \sum_n a_n \Psi_n , \quad \int [d\Phi] =
   \prod_n \mu_n (C_n) \int d a_n \quad ,
\label{norme}
\end{equation}
where $\mu$ is a normalization constant. In the generic case of higher-dimensional
quantum gravity there is little need to belabour the issue of what determines
the renormalization of $\mu$, since summing over the eigenvalue spectrum will give a
{\em divergent} result. This divergence needs to be regularized \cite{zeta}. Any such
regularization will introduce a scheme-dependent ambiguity in the normalization.
However, as we have emphasized above, this is not true in two-dimensional Weyl
invariant quantum gravity.

\vskip 0.1in
Following \cite{zeta}, we can write the functional determinant in
Eq.\ (\ref{eq:pathw}) in terms of a generalized zeta function as follows.
We begin with:
\begin{equation}
\int \prod_n \mu_n d a_n e^{-\half \lambda_n a_n^2 }
= \prod_n (2\pi)^{-1/2} \mu_n \lambda_n^{-1/2} =
     [{\rm det}(\half \mu^{-2} \pi^{-1} \Delta)]^{-1/2} \quad .
\label{normez}
\end{equation}
The infinite product can be rewritten as an infinite sum by taking the logarithm,
giving a formal expression on the right-hand-side that takes the form of
a generalized zeta function:
\begin{equation}
{\rm ln} [{\rm det}(\half \mu^{-2} \pi^{-1} \Delta)]^{-1/2}  =
 - \half (2\pi)^{-1/2} \sum_n \mu_n \log \lambda_n
 = \lim_{s\to 0} {{d}\over{ds}}
  \left [ (\half (2\pi)^{-1/2} )^{-s} \sum_n \mu_n \lambda^{-s}_n \right ]
 \quad .
\label{infsum}
\end{equation}
In the case of free embedding scalars, $\mu_n (C_n)$ is
independent of $n$, as was shown in \cite{poltorus}.\footnote{More
generally, the orthonormality constants, $C_n$, can depend on the
background fields of string theory. The normalization,
$\mu_n(C_n)$, turns out to be independent of $n$, as shown in
\cite{ncom}, where the precise form of the background field
dependence in an external two-form field has been derived.} Thus,
the normalization of the path integral for Weyl-invariant
two-dimensional quantum gravity is uniquely determined by the form
of the action, and by the gauge invariant measure for moduli. This
was shown clearly in Polchinski's 1986 derivation of the measure
for moduli in the one-loop bosonic string path integral
\cite{poltorus}, and in subsequent work on higher genus Riemann
surfaces including those with boundaries and crosscaps in
\cite{dhoker,polbook,mine,ncom}. The key point that remains is
explicit evaluation of the formally divergent right-hand-side of
this equation. Since the choice of worldsheet ultraviolet
regulator is unique, $\mu$ is unambiguously renormalized, and the
renormalization is, therefore, scheme-independent. The details of
such computations are reviewed in Appendix C using the contour
integral prescription given in \cite{zeta,poltorus,ncom}.

\vskip 0.1in Before leaving this general discussion, we should
emphasize that, although the quickest route to computing string
correlation functions employs conformal field theory techniques in
the operator formalism and, quicker still, operator product
expansions, these results are not unambiguously normalized. What
is unambiguous in operator formalism computations is the {\em
ratio} of two different correlation functions, and often, that
information suffices to describe all of the interesting physics.
This is precisely as in a generic quantum field theory: we have
not exploited the full power of the worldsheet formalism. However,
when we are interested in the numerical value of the vacuum energy
density and the cosmological constant per se, the physics is in
the string vacuum functional itself. We have no alternative but to
compute it from first principles, if possible, using the path
integral formalism. That such a calculation is viable in a
full-fledged ten-dimensional string theory as a consequence of its
relationship to {\em two-dimensional} quantum gravity, is nothing
short of a miracle. This is the significance of Polchinski's first
principles analysis of the Weyl-invariant Polyakov path integral
for critical string theory.\footnote{Many authors, including
myself \cite{mine}, are guilty of performing the apparent
sacrilege of introducing a non-Weyl invariant worldsheet regulator
into an analysis of the string theory path integral. In doing so,
the gauge invariance of the two-dimensional quantum gravity has
been explicitly broken. It should be emphasized, however, that for
the purposes of many calculations, such as those of asymptotic
bounds on the high energy string mass spectrum \cite{dhoker,mine},
the results are independent of the choice of worldsheet regulator.
This is {\em not} true of the computation of the normalization of
the vacuum energy where the use of a gauge invariant regulator is
crucial. Zeta-function regularization \cite{zeta} provides the
only correct answer. Note that in analyses where the worldsheet
gauge invariance has been explicitly broken, the string theory
path integral is primarily being invoked for its intuitive value
rather than as a high-precision computational tool.} As mentioned
in the Introduction, the result for the vacuum functional derived
in \cite{poltorus} is only the first step in deriving a number of
remarkable properties of the canonical ensemble of perturbative
strings, explored further in \cite{holo}. We should emphasize that
this approach can be adapted to any of the supersymmetric string
theories, as illustrated in detail for the interesting type I and
type I$^{\prime}$ unoriented, open and closed string theories in
appendices A thru C.

\vskip 0.1in A final comment on string theory correlation
functions. In quantum field theory we calculate what are called
normalized correlation functions by taking the {\em ratio} of an
$N$-point function divided by the vacuum functional, thus
eliminating the ambiguity introduced by a choice of ultraviolet
regulator as explained above. It should be emphasized that,
besides the vacuum amplitude at arbitrary loop order, the generic
loop correlation function in string theory is also unambiguously
normalized without the need to take a ratio \cite{ncom}. In
section 4, we will see an indication of this in the worldsheet
computation of the quantum of Dpbrane charge \cite{dbrane}. Since
the quantum of Dpbrane charge ia obtained by simply taking the
zero slope, or massless field theory, limit of the factorized
one-loop graph of the type I$^{\prime}$ string theory in the
background of a pair of Dpbranes, it is {\em required} to be
unambiguously normalized \cite{dbrane}. Thus, Dbrane charge
quantization simply follows as a consequence of the
normalizability of the vacuum amplitude in string theory
\cite{poltorus}.

\section{A Little Supergravity Background: Dpbrane Solitons}

It is helpful to begin our review of Polchinski's 1995 result for
the quantum of Dpbrane charge \cite{poltorus,dbr,dbrane} by
sketching the relevant insights from both the worldsheet and the
low-energy effective field-theory pictures that motivated this
calculation. The relationship of the worldsheet computation of the
normalized one-loop vacuum amplitude of type I string theory to
the quantum of Dpbrane charge will be described in the next
section. The low-energy effective field theory limit of the type
II closed string theories is a ten-dimensional $N$$=$$2$
supergravity theory without Yang-Mills gauge fields. The massless
supergravity Lagrangian can be extended by the inclusion of
kinetic and Chern-Simons terms for antisymmetric tensor fields,
$F_{p+2}$, corresponding to gauge potentials, $C_{p+1}$
\cite{teit,polbook}, where $p$ lies in the range, $-2$ $\le$ $p$
$\le$ $8$. This covers the full range of field strength tensors in
ten spacetime dimensions, namely, scalar to ten-form. Such a gauge
potential can couple to an extended object with a $p$$+$$1$
dimensional worldvolume, or $p$-brane, and it is natural to ask
whether the type II supergravities have classical solutions
describing dilaton-gravitational-antisymmetric-tensor field
configurations that have the geometry of a p-brane? The answer is
yes, and a large variety of such gravitational solitons have been
discovered over the years. The $p$-brane solitons of the type II
theories can be distinguished by whether they carry charge for a
Ramond-Ramond sector, or Neveu-Schwarz-Neveu-Schwarz sector,
antisymmetric tensor field strength \cite{pbrane}: the kinetic
term in the Lagrangian for the corresponding field strength
differs in the powers of $e^{-\phi}$ appearing in the pre-factor
\cite{witdual,polbook}.

\vskip 0.1in The next key point to note is that one of the
ten-dimensional supersymmetries is spontaneously broken by such a
choice of vacuum in either the type IIA or type IIB theory. We now
know that the corresponding ten-dimensional $N$$=$$1$
supergravities are the low-energy limits of the type I$^{\prime}$
and type IB string theories \cite{polbook}, an identification
originally made by Witten \cite{witdual} by direct comparison of
the ten-dimensional effective Lagrangians and spectrum of
low-lying masses. This also identified such vacua as BPS states of
the type II string theory. A further step was the recognition that
the p-brane solitons of the type II string theories could be
characterized by how their tension scales with the string coupling
\cite{witdual}: $1/g$ for R-R sector solitons vs $1/g^2$ for NS-NS
sector solitons. Shenker \cite{shenker} made the important
observation that R-R sector solitons would be responsible for
$e^{-1/g}$ corrections to the standard closed string perturbation
expansion in powers of $1/g^2$. This is unlike the NS-NS solitons
which give nonperturbative corrections of the form $e^{-1/g^2}$,
indistinguishable from those of an ordinary Yang-Mills gauge
theory soliton.

\vskip 0.1in All of these observations about the low-energy field
theory limit of string theory fall in place with Polchinski's
insight that Dirichlet-branes \cite{dbr}, suitably
supersymmetrized, are the carriers of R-R charge in the type II
string theories: in a vacuum that breaks half of the
supersymmetries and carries non-trivial R-R charge, the spectrum
of a type IIA or IIB closed string theory is extended by {\em an
open string sector with Dirichlet boundary conditions imposed on
the worldsheet fields} \cite{dbrane}. The key point is that the
closed string coupling scales as the {\em square} of the open
string coupling, and while the perturbation expansion of a pure
closed string theory closes on itself, it {\em might} allow
extension by an open string sector. Exceptions are the closed
heterotic string theories where the chiral worldsheet current
algebra prevents such an extension. However, this obstruction did
not exist for the type II string theories.

\vskip 0.1in The Dbrane vacua of the type II string theories could
thus be said to represent the more generic class of type II vacua,
having both an open and a closed string sector \cite{gp}. However,
since only half of the supersymmetries of the type IIA or type IIB
theories are preserved in such a vacuum, the Dbrane solutions
could equivalently be viewed as the classical vacua of an
$N$$=$$1$ ten-dimensional open and closed string theory. This
string theory is, respectively, either the type I$^{\prime}$ or
type IB string theory. As is well-known, type IIA and type IIB
were related by a T-duality transformation. This is also true for
type I$^{\prime}$ and type IB. The advantage of the latter
viewpoint is that questions which appeared obscure in the
Ramond-Neveu-Schwarz (RNS) worldsheet formalism of the type II
closed string theory, such as the prescription for Ramond-Ramond
vertex operators or the computation of the Ramond-Ramond sector
partition function, can now be straightforwardly answered in the
RNS formalism of the corresponding {\em open and closed} string
theories \cite{polbook}.

\vskip 0.1in The spacetime geometry of the type II vacuum carrying
pbrane charge is that of a $p$$+$$1$-dimensional hypersurface
embedded in ten-dimensional spacetime \cite{dbrane}. It is
well-known that a Hodge-star duality links a $p$-brane with a
($d$$-$$4$$-$$p$)-brane in $d$ dimensions, and that the
corresponding charges must satisfy a Dirac quantization condition
\cite{teit}. This is a simple consequence of applying quantum
mechanics to extended objects. Such objects were independently
discovered as the natural extension of the pointlike magnetic
monopoles of 4d gauge theory, by Savit, Orland, and Nepomechie
\cite{teit}. This early work on lattice field theories exploits
the well-known correspondence between the phase structure of
two-dimensional nonlinear sigma models and four-dimensional gauge
theories, and was followed by a more classical presentation of
higher pform gauge theories due to Teitelboim \cite{teit}. p-form
generalized electric-magnetic duality generalizes the
electric-magnetic duality of the abelian gauge field discovered by
Dirac. We can think of the magnetic monopole as a nonperturbative
configuration of the vector potential which couples to the
electron, the fundamental charge carrier of the electromagnetic
gauge field. Of course, a more careful analysis only finds
monopoles as stable configurations in field theories of scalars
coupled to Yang-Mills fields. It is the same for any pair of
pbranes that satisfy a Poincare duality relation \cite{polbook}:
\begin{equation}
\nu_p \nu_{d-4-p} = 2\pi n , \quad \quad n \in {\rm Z}  \quad ,
\label{eq:diracq}
\end{equation}
where $\nu_{d-4-p}$ is the quantized flux of the $(p+2)$-form
field strength. Such pbrane solitons are only found in
$d$-dimensional field theories with both scalars and antisymmetric
tensor fields of rank $p$$+$$2$, with $d$$\ge$$p$$+$$2$. In the
case of the Dirichlet p-brane solitons of the type I-I$^{\prime}$
string, where we have a clear understanding of the low-energy
effective field theory limit of an open and closed string theory,
we can show that {\em both} gravity and Yang-Mills gauge fields
exist in the worldvolume of the Dpbrane. This is because the
worldvolume represents the hypersurface in ten-dimensional
spacetime where open string end-points are permitted to lie, in
addition to the closed strings which, of course, lie in all ten
embedding spacetime dimensions. Thus, the scalar fields of the
ten-dimensional field theory split into worldvolume scalars and
bulk scalars, some of the latter representing the fluctuations of
the Dpbrane in the ten-dimensional embedding spacetime. Thus, in
the Dpbrane vacuum, the ten-dimensional Lagrangian obtained in the
low-energy limit of the relevant type II string theory acquires a
new term proportional to the Dpbrane worldvolume action. In the
physically relevant Einstein frame metric, related to the string
frame metric by a spacetime Weyl transformation:
$G_{\mu\nu}$$=$$e^{-4\Phi/(d-2)}G_{\mu\nu}^{\rm string}$, the
worldvolume action takes the simple form
\cite{dbr,dbrane,polbook}:
\begin{equation}
S_p = \tau_p \int d^{p+1} X e^{(p-3)\Phi/4}
  {\sqrt{-{\rm det} \left ( G_{\mu\nu} + B_{\mu\nu}
+ 2\pi\alpha^{\prime} F_{\mu\nu} \right )}} + \mu_{p} \int C_{p+1}
\quad , \label{eq:tens}
\end{equation}
where $\tau_p$ is the physical value of the Dpbrane tension.
$\mu_{6-p}$ is the quantum of magnetic D(6-p)brane charge, related
to the quantized flux of the field strength, $F_{p+2}$, obtained
by integrating it over a sphere in $(p+2)$ dimensions:
\begin{equation}
\int_{S_{p+2}} F_{p+2} = 2 \kappa_{10}^2 \mu_{6-p} \quad .
\label{eq:flux}
\end{equation}
Thus, the flux quantum satisfying a Hodge duality relation of the
form given in Eq.\ (\ref{eq:diracq}) is identified as, $\nu_p$
$\equiv$ $2 \kappa_{10}^2 \mu_p$. The flux quantum, $\nu_p$, and
the quantum of Dpbrane charge, $\mu_p$, differ from the
Dpbrane-tension, $\tau_p$, in their lack of dependence on the
closed string coupling. As a consequence, they can be calculated
unambiguously in weakly-coupled perturbative string theory.
Namely, we have the relations:
\begin{equation}
\kappa^2 \tau_p^2 \equiv g^2 \kappa_{10}^2 \tau_p^2 =
\kappa_{10}^2 \mu_p^2 , \quad\quad \kappa_{10}^2 \equiv \half
(2\pi)^7 \alpha^{\prime 4} \quad . \label{eq:charge}
\end{equation}
Here, $\kappa$ is the physical value of the ten-dimensional
gravitational coupling. and the dimensionless closed string
coupling, $g$ $=$ $e^{\Phi_0}$, where $\Phi_0$ is the vacuum
expectation value of the dilaton field.

\section{Worldsheet Computation of Dpbrane Charge}

\vskip 0.1in Given the remarkably simple worldsheet ansatz for
what are now called Dpbrane solitons of the type II string
theory--- pbrane solitons that carry charge under the $p$-form
gauge potentials of the R-R sector, and whose tension, $ \tau_p$,
scales as $1/g$, it is natural to ask whether the quantization of
Dpbrane charge can be inferred directly from a worldsheet
calculation? In fact, the result turns out to be stronger than
what one might have expected. Not only can Dpbrane charge,
$\mu_p$, be determined unambiguously in terms of the fundamental
string mass scale, $m_s$, but, upon substituting Polchinski's
result for $\nu_p$ \cite{dbrane} in the left hand side of Eq.\
(\ref{eq:diracq}), we find that the integer, $n$, on the right
hand side equals unity when $d$$=$$10$:
\begin{equation}
\nu_p^2 = 2\pi (4\pi^2 \alpha^{\prime})^{3-p} , \quad \quad \nu_p
\nu_{d-4-p} = 2\pi \quad {\rm (String ~ Theory)}  \quad .
\label{eq:diracqs}
\end{equation}
We emphasize that this string miracle has its origin in the
beautiful property of the one-loop vacuum amplitude of string
theory described earlier \cite{poltorus}: unlike the vacuum energy
density of a quantum field theory, the vacuum energy density of an
infra-red finite string theory is both ultraviolet finite, {\em
and normalizable}. Let us see how this works in detail.

\vskip 0.1in The tree-level Dpbrane tension is measured by the
tree-level coupling of the massless dilaton-graviton field to the
Dpbrane: in worldsheet language, this is the one-point function of
the massless dilaton-graviton closed string vertex operator on the
disk. Extracting the factor of $g$ by using the relation $\mu_p$
$=$ $g\tau_p$ gives the value of the quantum of Dpbrane charge,
$\mu_p$. It would be nice to determine this one-point function
directly from a first-principles path integral computation.
However, the normalization is tricky because of the dependence on
the volume of the conformal Killing group of the disk
\cite{grin,polbook}. In addition, one requires the result in the
supersymmetric type I string, and that computation has never been
done in the path integral formalism. In practice, it is much
easier to invoke the factorization of the annulus amplitude in
type I string theory to infer the normalization of the massless
closed string one-point function on the disk indirectly. This was
the method used to compute the tree-level Dpbrane tension in
\cite{dbrane}.\footnote{Polchinski's 1995 paper uses the operator
formalism, motivating the normalization of the amplitude rather
than deriving it. Although pedagogical, this presentation leaves
the origin of the lack of ambiguity in the quantum of Dpbrane
charge obscure. We will instead use the result of a path integral
derivation of the annulus amplitude following his 1986 paper
\cite{poltorus,mine}, reviewed in Appendix B.}

\vskip 0.1in
Referring to Eq.\ (\ref{eq:RRcharge}) in appendix B, we extract the contribution
from the annulus to the sum over connected worldsurfaces of vanishing Euler character,
$W_{\rm I-ann}$, in the background of a pair of parallel and static Dpbranes in the
type I string theory:
\begin{eqnarray}
 W_{\rm I-ann} &&=  \prod_{\mu=0}^{p} L^{\mu}
   \int_0^{\infty} {{dt}\over{2t}}
(8\pi^2 \alpha^{\prime}t)^{-(p+1)/2}
e^{-R^2 t /2\pi\alpha^{\prime}}
\nonumber\\
\quad && \quad \times
 {{1 }\over{\eta(it)^8 }} \left [
\left ({{ \Theta_{00} (0 , it)}\over{ \eta (it) }} \right )^4
- \left ( {{ \Theta_{0 1} (0 , it)}\over{ \eta (it) }} \right )^4
- \left ( {{ \Theta_{10} (0 , it)}\over{ \eta (it) }} \right )^4 \right ]
\quad .
\label{eq:pdbr}
\end{eqnarray}
The amplitude vanishes as a consequence of spacetime supersymmetry, as can be
seen by application of the abstruse identity for the Jacobi theta functions.
We will focus, therefore, on the contribution from massless spacetime bosons
alone, namely, the leading contributions from the $(00)$ and $(01)$ sectors.
The factorization limit corresponds to the long cylinder, $t$$\to$$0$. We
can expose the correct asymptotic behavior of the theta functions by expressing
them in terms of the theta functions with modular transformed argument,
$t$$\to$$1/t$:
\begin{eqnarray}
 W_{\rm I-ann} &&=  \prod_{\mu=0}^{p} L^{\mu}
   \int_0^{\infty} {{dt}\over{2t}}
(8\pi^2 \alpha^{\prime}t)^{-(p+1)/2}
e^{-R^2 t /2\pi\alpha^{\prime}}
\nonumber\\
\quad && \quad \times
 {{1 }\over{t^{-4}\eta(-1/it)^8 }} \left [
\left ({{ \Theta_{00} (0 , -1/it)}\over{ \eta (-1/it) }} \right )^4
- \left ( {{ \Theta_{0 1} (0 , -1/it)}\over{ \eta (-1/it) }} \right )^4
- \left ( {{ \Theta_{10} (0 , -1/it)}\over{ \eta (-1/it) }} \right )^4 \right ]
\quad .
\label{eq:pdbrm}
\end{eqnarray}
Expanding the theta functions in powers of $e^{-2\pi/t}$, and keeping only
the leading term in the expansion of the $(00)$ and $(01)$ sectors gives:
\begin{eqnarray}
 W_{\rm I-ann} &&\to
\prod_{\mu=0}^{p} L^{\mu}
   \int_0^{\infty} {{dt}\over{2t}}
(8\pi^2 \alpha^{\prime})^{-(p+1)/2}
e^{-R^2 t /2\pi\alpha^{\prime}} 2^{4} t^{(7-p)/2}
\nonumber\\
\quad &&= -
\prod_{\mu=0}^{p} L^{\mu} (8\pi^2 \alpha^{\prime})^{-(p+1)/2}
  2^{4} (2\pi\alpha^{\prime})^{(7-p)/2} \Gamma\left ({{7-p}\over{2}}\right ) |R|^{p-7}
\quad .
\label{eq:pdbrmf}
\end{eqnarray}
In the right-hand-side of this equality we can recognize the propagator
for a massless scalar field in $9$$-$$p$ dimensions, namely, in the bulk spacetime:
\begin{equation}
V_{p+1} (8\pi^2 \alpha^{\prime})^{-(p+1)/2}
  2^{4} (2\pi\alpha^{\prime})^{(7-p)/2} \Gamma \left ({{7-p}\over{2}}\right ) |R|^{p-7}
= 4\pi (4\pi^2 \alpha^{\prime})^{3-p} V_{p+1} G_{9-p}(|R|) \quad,
\label{eq:ident}
\end{equation}
Inverting the position space Green's function will enable us to make contact with
the field theory expression for the tree-level exchange of a graviton-dilaton
between two Dpbranes, the result for which depends explicitly on the Dpbrane tension.

\vskip 0.1in
The tree-level field theory calculation proceeds as follows. We begin with the
spacetime action in the Einstein frame metric given in Eqs.\ (\ref{eq:action}),
switching off both the ten-form, and the Yang-Mills, backgrounds.
The worldvolume action for the Dpbrane background takes the form given
in Eq.\ (\ref{eq:tens}).
Let us expand about the flat space background metric in perturbation theory,
$h_{\mu\nu}$$=$$G_{\mu\nu}$$-$$\eta_{\mu\nu}$, keeping terms upto
quadratic order in the field variations. We will perform this
calculation in the covariant Lorentz gauge:
\begin{equation}
F_{\nu} = \partial^{\mu} h_{\mu\nu} - \half
   \partial_{\nu} h^{\lambda}_{\lambda} = 0 \quad .
\label{eq:gaugeg}
\end{equation}
Thus, the gauge-fixed spacetime action takes the form:
\begin{eqnarray}
S^{\prime} [h,\Phi] =&& {{1}\over{2\kappa^2}}
\int d^{10} X {\sqrt{-G}}
  \left ( R_G - {{4}\over{d-2}}  (\partial \Phi)^2 \right ) -
 {{1}\over{4\kappa^2}} \int d^{10} X {\sqrt{-G}} F_{\nu} F^{\nu}
\nonumber\\
&&\quad\quad\quad
+ \tau_p \int d^{p+1} X e^{-(p-3)\Phi/4}  {\sqrt{-{\rm det} ~ G_{\mu\nu} }}
\nonumber\\
 =&& - {{1}\over{8\kappa^2}}
\int d^{10} X
  \left ( \partial_{\mu} h_{\nu\lambda} \partial^{\mu} h^{\nu\lambda}
  -  \half \partial_{\mu} h^{\nu}_{\nu} \partial^{\mu} h^{\lambda}_{\lambda}
  + {{16}\over{d-2}} \partial_{\mu} \Phi \partial^{\mu} \Phi \right )
\nonumber\\
&&\quad\quad\quad
- \tau_p \int d^{p+1} X \left ( {{p-3}\over{4}} \Phi -  \half h^{\lambda}_{\lambda}
        \right ) \quad .
\label{eq:field}
\end{eqnarray}
The Feynman graph of interest to us
in the diagrammatic expansion of this field theory
is the amplitude for the tree-level exchange of
the massless graviton-dilaton multiplet between a pair of Dpbranes in
$d$$=$$10$ spacetime dimensions.
Notice that both background, and propagators, for the massless dilaton
and graviton fields are decoupled when the action is written in the Einstein
frame metric. Thus, we need only
sum the corresponding tree-level Feynman graphs for each.
We will only need the trace-dependent piece of the graviton propagator
which mixes with the massless dilaton scalar exchange.
Given the action, Eq.\ (\ref{eq:field}), we can write down the form of
the free field propagators in the momentum space representation:
\begin{eqnarray}
<\Phi \Phi >  =&& - {{(d-2)\kappa^2}\over{4k^{2} }}
\nonumber\\
<h_{\mu\nu} h_{\sigma \rho}> =&&
 - {{2\kappa^2}\over{k^{2} }}
  \left ( \eta_{\mu\sigma} \eta_{\nu\rho} - \eta_{\mu\rho} \eta_{\nu\sigma}
- {{2}\over{d-2}} \eta_{\mu\nu}\eta_{\sigma\rho} \right )
\quad .
\label{eq:props}
\end{eqnarray}
Thus, for massless dilaton-graviton exchange in ten spacetime dimensions between
two Dpbranes, we have the tree-level Feynman amplitude:
\begin{eqnarray}
{\cal A} (k) =&&  {{(p-3)}\over{4}} \tau_p \cdot {{2\kappa^2}\over{k^{2} }}
\cdot  {{(p-3)}\over{4}} \tau_p
~+~ \half\tau_p \cdot {{2\kappa^2}\over{k^2}}
       \left ( 2(p+1) - {{1}\over{4}}(p+1)^2 \right ) \cdot \half\tau_p
\nonumber\\
=&& {{2\kappa^2}\over{k^{2} }}\tau_p^2
\quad .
\label{eq:amplr}
\end{eqnarray}
Recall that the momentum space representation of the Green's function, $1/k^{2}$,
is the Fourier inverse transform of the following position space Green's function:
  $G_{d}$$=$$ 2^{-2} \pi^{-d/2} \Gamma (\half d - 1) R^{2-d}$, a relation
that holds in any dimension $d$. Project $k$ on to a $(9$$-$$p)$-dimensional momentum vector
orthogonal to the worldvolume of the Dpbranes, and take the Fourier transform of
the result in Eq.\ (\ref{eq:amplr}) in $9$$-$$p$ space dimensions. Comparison with
Eq.\ (\ref{eq:pdbrmf}) gives the following result for $\tau_p$:
\begin{equation}
\nu_p^2 = 2 \tau_p^2 \kappa^2 = 2\pi (4\pi^2 \alpha^{\prime})^{3-p}
\quad .
\label{eq:identmu}
\end{equation}
Notice that the dependence on the dimensionless string coupling,
$g$, is eliminated in the product of physical parameters,
$\tau_p\kappa$, giving results for the flux quantum, $\nu_p$, and
the quantum of Dpbrane charge, $\mu_p$, that are valid even
outside of the realm of perturbation theory. This is as it should
be, not surprisingly, since $\nu_p$ is a spacetime topological
invariant which should always be computable, in principle, from
spacetime anomaly inflow arguments \cite{ghm}. Such computations,
carried out explicitly for the top-form field strength in
\cite{ghm}, have given independent confirmation of the correctness
of Polchinski's worldsheet interpretation of the carriers of R-R
charge \cite{dbrane}.

\section{Normalization of Worldvolume EFT Couplings}

\vskip 0.1in Consider a background of the type I or type
I$^{\prime}$ string theory with given set of Dbranes. The
spacetime geometry and background fields and fluxes on the
worldvolume of the Dbranes, or intersections thereof, constitute a
{\em braneworld}, with associated worldvolume effective field
theory Lagrangian. Let us explain why it is possible to determine
{\em all} of the couplings in such a worldvolume Lagrangian,
inclusive of numerical factors, in terms of the single mass scale,
$m_s$, plus a finite number of dimensionless scalar field
expectation values, or moduli. Our observation exploits a property
of the annulus graph of the open and closed string theory first
noted by us in \cite{ncom}: in order to extract the contribution
from the lowest-lying open string modes dominating the short
cylinder limit of a given scattering amplitude, there is no need
to set the modulus of the annulus to some ad-hoc value, thereby
explicitly breaking reparametrization invariance. Instead, we will
first expose the asymptotics of the integrand by Taylor expansion,
followed by an explicit evaluation of the modular integral. The
physical significance of having thereby preserved all of the
manifest symmetries of the worldsheet formalism, namely, exact
Weyl and diffeomorphism invariance, is that we obtain an {\em
unambiguously normalized} expression for each of the couplings in
the worldvolume effective field theory.

\vskip 0.07in We begin with some general statements on loop
correlation functions in string theory. The full beauty of
perturbative string theory becomes transparent upon detailed
examination of the loop expansion. The world-sheet representation
of string loop amplitudes implies the existence of a single graph
at each order in loop perturbation theory. In precise analogy with
the vacuum functional, the scattering amplitudes of string theory
can be equivalently interpreted as quantum correlators in a
two-dimensional \lq\lq gauge" theory, where the gauge symmetry in
question is two-dimensional diffeomorphism and Weyl invariance
\cite{polyakov}. The S-matrix describing the scattering of
asymptotic string states is obtained by invoking two-dimensional
conformal invariance to represent asymptotic on-shell states as
operator insertions that are local in the {\em two-dimensional}
sense, i.e., localized on the worldsheet. Off-shell string states
are macroscopic boundaries of the worldsheet--- either macroscopic
closed loops or macroscopic line segments, localized instead in
the embedding {\em spacetime} \cite{cmnp,gr,wils}. This
correspondence between worldsheet and spacetime pictures has
remarkable consequences. Consider a gauge invariant path integral
expression for the generic Greens function at arbitrary order in
the string loop expansion. Upon gauge fixing to conformal gauge,
the path integral over metrics is restricted to the fiducial
representative from each conformally inequivalent class of
metrics. This is an {\em ordinary} integral over the finite number
of moduli parameters of Riemann surfaces of fixed topology.
Remarkably, in any critical string theory, both the measure of the
path integral, the functional determinants, and vertex operator
insertions, can be unambiguously computed as functions of the
moduli at least in principle, while preserving the full
Diffeomorphism $\times$ Weyl gauge invariance. The result is an
unambiguously normalized and ultraviolet finite expression for
both the Greens function and generic N-point interactions, with a
well-defined, zero string tension, field theory limit.

\vskip 0.07in The resulting expressions for the field theory
Greens functions are free of ultraviolet regulator ambiguity. More
importantly, in an infrared finite string theory, they are also
free of ambiguity in the choice of renormalization scheme
\cite{poltorus,ncom}. Both properties are a consequence of having
maintained manifest two dimensional general coordinate invariance
in computing the full string theory Greens function {\em prior to
taking the field theory limit} which we define as follows: we
factorize on massless mode exchange in either open or closed
string sectors, projecting also onto the massless on- or off-shell
modes in any external states, and integrating out the worldsheet
modulus dependence of the resulting expression. The result is a
coupling in the field theory in which we can smoothly take the
zero string tension limit. Thus, the expression for any
renormalized string theory N-point Greens function including, in
particular, the corresponding field theory limit, is independent
of dependence on the string tension, $m_s$ $\sim$ $\alpha^{\prime
-1/2}$, which plays the role of an ultraviolet cutoff in
spacetime.

\vskip 0.1in As an illustration of our comments, let us quote two
results from the work in Ref.\ \cite{ncom}, done in collaboration
with Novak. This describes a first-principles covariant path
integral derivation of the $N$-point one-loop scattering
amplitudes of open and closed bosonic string theory in an external
two-form background field of generic strength. This is a
straightforward adaptation of the derivation of the N-point
scattering amplitude of the closed bosonic string given in
\cite{poltorus}. The main new subtleties in the zeta function
regularization technique resolved in \cite{ncom} have to do with
the extension to both planar and nonplanar amplitudes, and as a
consequence of the external field dependence in the expressions.
The results provide confirmation of the basic insights into
renormalization given by the path integral approach
\cite{polyakov,poltorus}, in the framework of an open and closed
string theory with nontrivial backgrounds. Evaluating the Greens
function for two open string tachyon vertex operators on the same
boundary of the annulus in an external twoform background of
generic strength \cite{ncom}:
\begin{eqnarray}
G^{\prime} (\sigma_i,\sigma_j)  =&& - {{2 \alpha^{\prime} }\over{1
+ B^2}}
  {\rm ln} {{|\Theta_{1}({{it \sigma_{ij}}\over{2}}, {{it}\over{2}}
     ) |^2}\over{ |\eta( {{it}\over{2}} ) |^2 }}
+ {{2 \pi \alpha^{\prime} t}\over{1+B^2}}
   \left[(\sigma_{ij})^2 - {{2B\sigma_{ij}}\over{t}} +{{2B^2}\over{3t^2}}
\right ]
 + {{ 2 \pi \alpha^{\prime} B}\over{1+B^2}}
 {{\sigma_{ij}}\over{|\sigma_{ij}|}} \quad . \label{eq:grbndry}
\end{eqnarray}
Note that the short distance divergence on the worldsheet can be
extracted by subtractive renormalization while preserving
two-dimensional reparametrization invariance: we express the
subtraction in terms of the geodesic distance between the two
sources on the worldsheet \cite{poltorus,dhoker,ncom}. The short
distance limit gives:
\begin{equation}
G^{\prime} (\sigma_i , \sigma_i ) = - {{ 2 \alpha^{\prime}
}\over{1 + B^2}} ~ {\rm ln} ~ | t \sigma_{ij} |^2 - {{ 2
\alpha^{\prime} }\over{1 + B^2}} ~ {\rm ln} ~ | \pi
(\eta(it))^2|^2 + {{4 \pi \alpha^{\prime} B^2}\over{3t(1+B^2)}} ,
\quad \sigma^2_i = 0, ~ 1
 \quad .
\label{eq:distinv}
\end{equation}
and we can define the renormalized Greens function:
\begin{equation}
\lim_{\sigma_i \to \sigma_j} G^{\prime}_P (\sigma_i , \sigma_j ) =
- 2 \alpha^{\prime} ~ {\rm ln} ~ d^2 (|\sigma_{ij}|)
 + f(\sigma_i, \sigma_j) \quad ,
\label{eq:dist}
\end{equation}
where $d$ is the distance between the sources as measured on the
world-sheet, and the function $f$ is finite in the limit
$\sigma_i$$=$$\sigma_j$. Notice that the leading short distance
divergence has the same form as on the boundary of the disk
\cite{polbook}, except that it is defined with respect to the
fiducial metric on the annulus. From the viewpoint of the
correlation function itself, the divergent terms should be
understood as having been absorbed in a renormalization of the
bare open string coupling \cite{poltorus,ncom}. The precise
expression for the one-loop, N-point, planar scattering amplitude
can be found in \cite{ncom}; we will extract only its low energy
field theory limit by exposing the large $t$ asymptotics of the
short cylinder limit. A full discussion of both planar and
nonplanar amplitudes, with external field dependence intact, can
be found in \cite{ncom}. For clarity, we consider the massless
limit of the planar amplitude, with all $N$ on-shell vertex
operators on the Dpbrane, and with the background field set to
zero:
\begin{eqnarray}
{\cal A}_P |_{{\rm massless}} &&=
  i g^N \delta ( \sum_{i=1}^{N} {\bf p}_i )
 \left [ \prod_{r=1}^{N} \int
d \sigma_r \right ]
  \int_0^{\infty} {{dt}\over{2t}} (2t)^{N-(p+1)/2}
 \nonumber \\
&& \quad \quad \times \prod_{i \neq j;~ i,j=1}^{N}
     \left [  e^{ - \pi t \sigma_{ij}^2 {\bf k}_i \cdot  {\bf k}_{j} }
~ | {{2}\over{\pi}} {\rm Sinh}( \half \pi t
   \sigma_{ij})|^{ 2 {\bf k}_i \cdot {\bf k}_{j} }
 ~ \left ( 24 - 8 {\bf k}_i \cdot
{\bf k}_{j} e^{-\pi t \sigma_{ij}}  \right ) + O(e^{-2\pi t})
\right ] \cr && \label{eq:formprlim}
\end{eqnarray}
Notice that the integral over the cylinder modulus can be
performed explicitly, giving a closed form expression for the
$N$-point function in the worldvolume gauge theory: the Feynman
graph is the circle with $N$ external legs, and a single photon
running around the loop. It should be emphasized that the
corrections from massive open string states circulating in the
loop do not change the basic form of this result, the single term
given here being replaced by a series of the form:
\begin{equation}
\sum_{n=0}^{\infty} ~ F^{({\rm open})}_n \left ( {\rm
Sinh}(\half\pi t \sigma_{ij}) \right )
    e^{-(\sigma_{ij}+n)\pi t}
\quad , \label{eq:poly}
\end{equation}
where $F^{({\rm open})}_n$ is a polynomial function of the ${\rm
Sinh}(\half\pi t \sigma_{ij}) $, $i$,$j$$=$$1$, $\cdots$, $N$.

\vskip 0.1in A more realistic calculation requires extension to a
stack of coincident Dpbranes, with specific nonabelian gauge group
realized on them. The $N$ open string tachyon vertex operator
extensions would be replaced by $N$ massless gluon vertex
operators. Finally, this particular computation would be performed
for the open string sector of a fully consistent, anomaly-free,
type I or type I$^{\prime}$ string theory ground state. None of
these features brings in insurmountable difficulty: the requisite
worldsheet technology is well-developed, and can be found in the
standard sources. An interesting observation can be made about
bulk-boundary couplings by considering the opposite limit of the
planar amplitude, namely, the long cylinder limit. Factorization
of the generic $K$-point planar correlation function on the
annulus will yield the normalized $(K,1)$-point function on the
disk, with $K$ open string, and one massless closed string, vertex
operator insertions. The field theory limit of this computation
gives the {\em tree-level} coupling of an arbitrary number of
worldvolume gauge bosons to the bulk dilaton-graviton. Notice, in
particular, that the worldvolume Yang-Mills gauge theory does {\em
not} decouple from the bulk supergravity theory, even at lowest
order in the Yang-Mills gauge coupling.

\section{The Macroscopic Loop Amplitude}

\vskip 0.1in Let us return to the familiar computation of the
contribution to the one-loop amplitude of bosonic open and closed
string theory from world-surfaces with the topology of a cylinder,
reviewed from first principles in the covariant path integral
approach in appendix A.2. From the perspective of the closed
string channel, this graph represents the tree-level propagation
of a single closed string, exchanged between a spatially-separated
pair of Dpbranes. A crucial observation is as follows: although
the Dpbrane vacuum corresponds to a spontaneous breaking of
translation invariance in the bulk $25$$-$$p$ dimensional space
orthogonal to the pair of Dpbranes, notice that spacetime
translational invariance is preserved within the
$p$$+$$1$-dimensional worldvolume of each Dpbrane.

\vskip 0.1in It is interesting to ask whether it is possible to
modify this calculation such that {\em all} $26$ spacetime
translation invariances are broken. We emphasize that we ask this
question not only for the point-like boundary limit of the annulus
graph, but for the annulus with {\em macroscopic} boundary loops.
The former limit with pointlike boundaries corresponds to the
tree-level exchange of a closed string between a pair of
Dinstantons: their worldvolumes are spacetime points, and each
boundary of the annulus is therefore mapped to a point in the
embedding $26$d spacetime. The latter case corresponds to a
genuinely new worldsheet amplitude, and the corresponding analysis
of the covariant string path integral brings in many new features
first described in \cite{cmnp,wils}, and reviewed in appendix D of
this paper. Remarkably, we will find that this computation leads
us to discover a new Dirichlet vacuum of open and closed bosonic
string theory, namely, that of a Dirichlet (-2)brane. This
identification was first made by Chaudhuri in \cite{flux},
examining evidence from both the target spacetime and worldsheet
perspectives of the type I$^{\prime}$ string. A complementary
interpretation from the perspective of the boundary state
formalism appears in our recent work \cite{hodge}; we also give a
discussion of the relation to the (-1)form potential in the global
symmetry algebra of the massive type IIA supergravity \cite{sw}.

\vskip 0.1in It is convenient to align the macroscopic loops,
${\cal C}_i$, ${\cal C}_f$, which we will choose to have the
common length $L$, such that their distance of nearest separation,
$R$, is parallel to a spatial coordinate, call it $X^{25}$. As in
appendix A.2, the Polyakov action contributes a classical piece
corresponding to the saddle-point of the quantum path integral:
the saddle-point is determined by the minimum action worldsurface
spanning the given loops ${\cal C}_i$, ${\cal C}_f$. The result
for a generic classical solution of the Polyakov action was given
by Cohen, Moore, Nelson, and Polchinski in Ref.\ \cite{cmnp}. For
coaxial circular loops in a flat spacetime geometry, we have a
result identical to that which holds for a spatially separated
pair of generic Dpbranes in flat spacetime \cite{wils}.

\vskip 0.1in The details of the computation are reviewed in
appendix D. The main difference from the analysis of the annulus
diagram of open and closed string theory is the implementation of
boundary reparametrization invariance: although all 26 coordinates
of the embedding worldvolume of the space-filling D25brane are
chosen to satisfy the Dirichlet boundary condition,
reparametrization invariance of the map from the boundaries of the
annulus to a given pair of loops in spacetime requires that we
allow nontrivial boundary reparametrizations of the worldsheet
metric (einbein) on the boundary \cite{cmnp,wils}. The result is
an additional contribution to the measure for moduli in the path
integral, computible in terms of the functional determinant of the
scalar Laplacian on the one-dimensional boundary of the
worldsheet. Our result for the connected sum over worldsurfaces
with the topology of an annulus with boundaries mapped onto
spatially separated macroscopic loops, ${\cal C}_i$, ${\cal C}_f$,
of common length $L$ takes the form \cite{cmnp,wils}:
\begin{equation}
 {\cal A} = i \left [ L^{-1}(4\pi^2 \alpha^{\prime})^{1/2}
 \right ]
   \int_0^{\infty} {{dt}\over{2t}} \cdot (2t)^{1/2} \cdot
    \eta (it )^{-24} e^{-R^2 t /2\pi\alpha^{\prime}}
\quad . \label{eq:resulttafb}
\end{equation}
The only change in the measure for moduli is the additional factor
of $(2t)^{1/2}$ contributed by the functional determinant of
${\cal J}$. The pre-factor in square brackets is of interest;
recall that there is no spacetime volume dependence in this
amplitude since we have broken translational invariance in all
$26$ directions of the embedding spacetime. If we were only
interested in the point-like off-shell closed string propagator,
as in \cite{cmnp}, the result as derived is correct without any
need for a pre-factor.\footnote{Comparing with the final
expression for the off-shell point-like propagator given in Eq.\
(4.5) of \cite{cmnp}, and setting $t$$\to$$2\lambda$ in order to
match with the notation in \cite{cmnp}, the reader should ignore
an extraneous factor of $\lambda^{-13}$, which should clearly be
absent in an all-Dirichlet string amplitude.} However, we have
{\em required} that the boundaries of the annulus are mapped to
loops in the embedding spacetime of an, a priori, fixed length
$L$. Since a translation of the boundaries in the direction of
spacetime parallel to the loops is equivalent to a boundary
diffeomorphism, we must divide by the (dimensionless) factor: $L
(4\pi^2 \alpha^{\prime})^{-1/2}$. This accounts for the pre-factor
present in our final result. Note that for more complicated loop
geometries, including the possibility of loops with corners, the
pre-factor in this expression will take a more complicated form.

\vskip 0.1in As mentioned above, we suspect that this expression
can be interpreted as computation of the one-loop vacuum amplitude
in a {\em distinct} Dirichlet background of the familiar open and
closed string theory in critical spacetime dimension. As a check,
let us take the factorization limit of the amplitude, expressing
the eta function in an expansion in powers of $q$$=$$e^{-2\pi/t}$.
The small $t$ limit is dominated by the lowest-lying closed string
modes and the result is:
\begin{eqnarray}
 {\cal A} =&& i \left [ L^{-1}(4\pi^2 \alpha^{\prime})^{1/2}
 \right ]
   \int_0^{\infty} dt\cdot (2t)^{-1/2} \cdot t^{12} \cdot q^{-1}
    \left ( 1 + 24 q + O(q^2) \right ) e^{-R^2 t
    /2\pi\alpha^{\prime}}\cr
\to&& i L^{-1}
 24 \cdot 2^{-12}(4\pi^2 \alpha^{\prime})^{13} \pi^{-25/2} \Gamma \left ( {{25}\over{2}} \right )
 |R|^{-25} \quad .
\label{eq:resulttafbt}
\end{eqnarray}
Repeating the steps reviewed in section 4 \cite{dbrane,polbook},
we infer the existence of a Dirichlet (-2)brane in open and closed
bosonic string theory with tension:
\begin{equation}
\tau_{-2}^2 = {{\pi}\over{256 \kappa^2}} (4\pi^2
\alpha^{\prime})^{13} \quad . \label{eq:tension-2}
\end{equation}
How does one give physical meaning to a Dirichlet (-2)brane? In
order to understand its origin in the spectrum of pbranes, it will
be helpful to consider the meaning of a D(-2)brane from the
perspective of both the boundary state formalism, and of the
ten-dimensional supergravity dualities.

\vskip 0.1in Let us begin from the perspective of the boundary
state formalism, originally developed by Callan, Lovelace, Nappi,
and Yost in  \cite{call}, assuming open strings with Neumann
boundary conditions, and later adapted to the Dirichlet case by
Green \cite{gr}. The extension to the boundary state of a generic
Dpbrane in the superstring is due to Li \cite{li}. The boundary
state is simply the physical state in the Hilbert space that is
annihilated by the boundary conditions on all two-dimensional
fields, when interpreted as operator statements \cite{call}. We
remind the reader that the worldsheet metric of the classical
Polyakov action is eliminated in favor of the Faddeev-Popov
$b$,$c$, ghost fields \cite{polbook} in the course of covariant
quantization. Thus, in bosonic string theory, all we have left are
the Dirichlet boundary conditions acting on the 26 two-dimensional
free scalar fields, and corresponding conditions on the ghosts. In
terms of the free bosonic oscillators, the Dinstanton boundary
state therefore satisfies the constraint \cite{gr}:
\begin{equation}
(\alpha_n^{\mu} - \alpha_{-n}^{\mu} ) |B, y^{\mu} > =0 , ~
[\alpha_m^{\mu} , \alpha_n^{\nu} ] = m \delta_{mn}
\delta^{\mu\nu}, \quad  \forall \mu=0, \cdots , 25 \quad .
\label{eq:diri}
\end{equation}
For the $b$, $c$ ghosts, we have corresponding constraints:
\begin{equation}
(b_n - {\tilde{b}}_{-n} ) |B, y^{\mu} > =0 , \quad (c_n +
{\tilde{c}}_{-n} ) |B, y^{\mu} > =0 , \quad [ b_n , c_m ] =
\delta_{m+n} = [ {\tilde{b}}_n , {\tilde{c}}_m ] \quad .
\label{eq:dirig}
\end{equation}
Since the boundary state is a physical state in the Hilbert space,
$|B>$ is annihilated by the BRST charge, the operator responsible
for implementing two-dimensional reparametrization invariance. In
particular, $|B>$ is also {\em required} to be invariant under
diffeomorphisms of the boundary: $\sigma^2 $ $\to$ $f(\sigma^2)$,
of the worldsheet annulus. This last constraint is trivially
satisfied by the Dirichlet boundary states, $-1$ $\le$ $p$ $\le$
$25$, since all two-dimensional fields vanish on the boundary. In
the extreme case of the Dinstanton, where the worldvolume is
simply a spacetime point, the map from the boundary of the annulus
to the worldvolume of the Dinstanton is trivial. The result takes
the form \cite{gr}:
\begin{equation}
|B> = \exp \left [ \sum_{n=1}^{\infty} \left ( \alpha^{\mu}_{-n}
{\tilde{\alpha}}^{\mu}_{-n} - b_{-n} {\tilde{c}}_{-n} -
{\tilde{b}}_{-n} c_{-n} \right ) \right ] | \Omega > \quad .
\label{eq:dirigs}
\end{equation}
Since we have assumed rigid Dirichlet boundaries, $|\Omega>$ is
the $SL(2,{\rm C})$ invariant vacuum with zero (transverse)
momentum orthogonal to the boundary \cite{polbook}.

\vskip 0.1in The boundary conditions we have imposed in our
calculation of the macroscopic loop amplitude do not permit a
trivial resolution to this last constraint. Requiring that the
boundaries of the annulus map into {\em loops at fixed locations}
in the embedding spacetime implies that we have broken all 26
spacetime translations. Thus, it is helpful to begin with the
Dinstanton boundary state described above. Next, we must insert an
operator in the transfer matrix whose purpose is to smear the
location of the Dirichlet end-point over the specified loop ${\cal
C}_i$, and likewise for ${\cal C}_f$. It is convenient to identify
the direction parallel to the pair of coaxial circular loops as
one of the coordinates of spacetime, call it $X^i$. It follows
that the operator that must be inserted in the transfer matrix is
none other than the {\em momentum} generator, ${\hat{P}}_i$. Thus,
we conclude that the boundary state that describes a
Dirichlet(-2)brane is obtained by simply acting with the momentum
generator on the boundary state of the Dinstanton.

\vskip 0.1in Is it possible to make this identification more
precise? We remind the reader that the normalization of a boundary
state is, a priori, ambiguous, as already noted in \cite{call,gr}.
We should point out, however, that the normalization of the
boundary state could be determined by matching to the
factorization limit of the covariant path integral computation: as
shown in section 4, and above, the one-loop amplitude factorizes
on a closed string propagator connecting (normalized) one-point
functions on the disk, each localized on a Dpbrane.

\vskip 0.1in The corresponding one-loop amplitudes in the operator
formalism can be written as follows. Identifying the modular
parameter of the one-loop amplitude as a fictitious \lq\lq time",
the integrand can be interpreted as the transfer matrix for
two-dimensional field theory data from an initial boundary state,
$|B_i>$, to a final state, $|B_f>$. The precise measure of the
one-loop string amplitude is difficult to motivate from first
principles without the path integral derivation, but the integral
over $t$ is easily justified by invoking two-dimensional conformal
invariance. Thus, the sum over world-surfaces with the topology of
an annulus, and with boundaries terminating on a spatially
separated pair of Dinstantons, or D(-1)branes, can be equivalently
interpreted in terms of the transfer matrix formalism as follows
\cite{call,li}:
\begin{equation}
 {\cal A}_{-1} (R) =
   \int_0^{\infty} dt <B,R| e^{- (L_0 + {\tilde{L}}_0 - 1)t } | B,0>
    e^{-R^2 t /2\pi\alpha^{\prime}}
\quad , \label{eq:resultr}
\end{equation}
where $L_0$, ${\tilde{L}}_0$, are the zero modes of the Virasoro
generators, including both matter and ghost degrees of freedom,
the vacuum energy of the $SL(2,{\rm C})$ invariant vacuum is $-1$,
and we have separated the classical contribution to the
Hamiltonian. Comparing with the expressions in appendix A.2, and
in section 4, the reader will note that the normalization of the
Dinstanton boundary state and, consequently, of the one-loop
string amplitude is, apriori, unknown and ambiguous
\cite{call,gr}. The corresponding expression for the channel
amplitude in the case of the Dirichlet(-2)branes takes the form:
\begin{equation}
 {\cal A}_{-2} (R) =
   \int_0^{\infty} dt <B,R|{\hat{P}}^{\dagger} e^{- (L_0 + {\tilde{L}}_0 - 1)t } {\hat{P}} | B,0>
    e^{-R^2 t /2\pi\alpha^{\prime}}
\quad , \label{eq:resultr2}
\end{equation}
where ${\hat{P}}$ acts in the direction parallel to the coaxial
circular loops. The path integral expression derived in appendix D
gives concrete meaning to the various terms in the operator
representation.

\vskip 0.1in Additional insight into the nature of the D(-2)brane
is provided by supergravity considerations. Recall that the
ten-form field strength is the Hodge dual of a scalar field
strength in ten spacetime dimensions. Indeed, Roman's massive IIA
supergravity theory is known to have a nine-form potential which
couples to the D8brane of the IIA string theory
\cite{romans,dbrane}. Let us write down the relevant equations for
the coupling to a D8brane in the ten-form formulation of the
massive type IIA supergravity theory. In the Einstein frame
metric, the bosonic part of the massive IIA action takes the
simple form \cite{polbook}:
\begin{equation}
S = {{1}\over{2\kappa_{10}^2}} \int d^{10} X {\sqrt{-G}}
  \left ( R_G - {{4}\over{d-2}} (\partial \Phi)^2  - \half e^{5\Phi/2} M^2 \right )
+  {{1}\over{2\kappa_{10}^2}} \int M F_{10}  \quad ,
\label{eq:action}
\end{equation}
where $M$ is an auxiliary field which will be eliminated by its
equation of motion. $F_{10}$ is the non-dynamical ten-form field
strength, which can be dualized to a zero-form, or scalar, field
strength, $*F_{10}$. This non-dynamical constant field generates a
uniform vacuum energy density that permeates the ten-dimensional
spacetime, thus behaving like a cosmological constant: $F_{10}$
$=$ $d C_9$, varying with respect to the gauge potential, $C_9$,
gives $M$ $=$ constant, and varying with respect to $M$ gives,
$F_{10}$ $=$ $ M e^{5\Phi/2} V_{10}$, where $V_{10}$ is the volume
of spacetime. Thus, we can identify the dualized scalar field
strength: $*F_{10}$ $=$ $M e^{5\Phi/4}$. Notice that, since the
objects appearing explicitly in the spacetime Lagrangian are the
field strengths, there is, of course, nothing puzzling about the
existence of a scalar field strength in the Ramond-Ramond sector
of the type IIA string theory. The Dirichlet (-2)brane should be
identified as the source for the associated topological charge.

\section{Conclusions}

\vskip 0.07in A basic insight that follows from the observations
summarized in this paper is that perturbative string theory is
more simply understood as an exactly renormalizable
two-dimensional field theory with precisely {\em one} independent
Wilsonian renormalized parameter \cite{ncom}. This insight is
already apparent in the computation of the one-loop
renormalization of the string coupling in closed bosonic string
theory by Polchinski in \cite{poltorus}. We have added further
clarifications, especially in the interesting case of the open and
closed type I and type I$^{\prime}$ string theories in generic
twoform background \cite{ncom,holo}.

\vskip 0.07in It should be noted that the exact renormalizability
of perturbative string theory is obscure from the viewpoint of the
spacetime effective Lagrangian: computation of the renormalized
Greens functions for massless fields in an $\alpha^{\prime}$
expansion, while keeping only a finite number of terms in the
nonrenormalizable Wilsonian effective action, does not enable one
to infer that the ultraviolet cutoff can, in fact, be removed.
Thus, at any finite order in the $\alpha^{\prime}$ expansion,
string theory is indeed nonrenormalizable. Our assertion that
perturbative string theory is exactly renormalizable relies
crucially on knowledge of an equivalent, {\em all}-orders in
$\alpha^{\prime}$, worldsheet representation of the Greens
function and loop correlation functions. It should be emphasized
that such an analysis can only be carried out in explicit form in
the flat spacetime background, known to be infra-red finite to all
orders in string perturbation theory for an anomaly-free choice of
nonabelian gauge group \cite{holo}.

\vskip 0.1in Perhaps the most important issues raised by this
paper are the observations in section 5, where we have noted that
the remarkable normalizability property of the perturbative string
scattering amplitudes can survive in the low energy effective
field theory limit: all of the couplings in the worldvolume
effective Lagrangian, inclusive of numerical factors, can be
determined in terms of the single mass scale, $m_s$, plus a finite
number of dimensionless scalar field expectation values, or moduli
\cite{ncom}. As mentioned earlier, this could have significant
consequences for the derivation of precision perturbative gauge
theory results from perturbative string scattering amplitude
computations. This is an area of considerable import for the
continued development of high energy physics, both in precision
particle physics and astrophysics, and for braneworld cosmologies.
It appears to us very unfortunate that the full power of
perturbative string theory computations has not been fully
exploited in its most immediate application to low energy physics:
namely, the direct derivation of scattering amplitudes in
perturbative gauge theories.

\vskip 0.1in The physical significance of the Dirichlet(-2)brane
described in section 6, and the implications for Poincare duality
in String/M theory, are discussed elsewhere \cite{hodge}. As
mentioned earlier, macroscopic loop amplitudes in string theory
have the potential to provide many new applications to physics.
The preliminary analysis of loop geometries given in
\cite{cmnp,wils}, including the interesting possibility of {\em
corners}, is deserving of further development. There exist a
number of potential applications relating to cosmic string
production from the vacuum, cosmic string scattering, as well as
the study of radiation from cusps and/or corners on a cosmic
string. It would be remarkable to have a first principles
formalism for deriving such phenomena directly from a fundamental,
and fully renormalizable, quantum theory.

\vskip 0.2in \centerline{\bf ACKNOWLEDGEMENTS}

\medskip
The notes that appear in Appendix C on the zeta function
regularization of divergent sums typically appearing in one-loop
string amplitudes were written up in the Spring '03, during the
course of an email collaboration with Kristj\'an Kristj\'ansson
and Lar\'us Thorlacius. I thank them for helpful input on the
presentation of the derivations. A concise presentation of this
technique, but with more sophisticated applications, can be found
in my paper with Eric Novak \cite{ncom}. I thank Peter Orland for
bringing the work in Ref [15] to my notice.

\appendix

\section{Open and Closed Bosonic String Theory}

\vskip 0.1in Consider a pair of parallel Dpbranes separated by a
distance $R$ in the direction $X^{p+1}$ with $p$ $<$ $25$. There
are four worldsheet diagrams that contribute to the one-loop
amplitude in an unoriented open and closed string theory. Recall
that worldsheets with Euler number, $\chi$ $=$ $2$ - $2h$- $b$ -
$c$, where $h$, $b$, and $c$, are, respectively, the number of
handles, boundaries, and crosscaps on the worldsheet, contribute
at $O(g^\chi)$ to the string perturbative expansion
\cite{polbook}. We will perform the sum over surfaces for each of
these topologies in turn, starting with the closed bosonic string
amplitude derived in \cite{poltorus}.

\subsection{One-loop Vacuum Amplitude: Torus}

This is essentially a review of Polchinski's 1986 derivation of
the one-loop vacuum amplitude for closed bosonic string
theory.\footnote{Sections in Polchinski's text that may be helpful
in reading this appendix are 3.2, 3.5, 5.1, 5.2, 7, 10.8, and
13.5. The presentation here is based on his 1986 paper
\cite{poltorus}, and the follow-up works \cite{dhoker, mine},
rather than on the textbook. One reason is that we wish to
highlight the relationship to path integrals in quantum gravity,
eschewing the use of worldsheet holomorphicity and ghosts,
although, as we have emphasized earlier, conformal field theory in
the operator formalism has its uses. More importantly, the casual
reader of Polchinski's text will fail to notice that many results,
such as Eqs.\ (7.2.3), (7.2.4), and all of secs.\ 7.4, 10.8, and
13.5, cannot be obtained in closed form without the
zeta-regularized path integral derivation. The author has done a
nice job of {\em motivating} the results but, in the interests of
pedagogy, many significant derivations have been skipped.} Since
closed strings cannot couple to a background magnetic field, the
background field strength, and Dpbrane geometry, are of no
relevance to the computation of this diagram. Note that the
worldsheets of toroidal topology are embedded in all 26 spacetime
dimensions.

\vskip 0.1in
The generic metric on a torus can be parameterized by two shape parameters,
or worldsheet moduli, $\tau$$=$$\tau_1 $$+$$i \tau_2$, and it takes the form:
\begin{equation}
ds^2 = e^{\phi} |d\sigma^1 + \tau d \sigma^2|^2 , \quad {\sqrt{g}} = e^{\phi} \tau_2
, \quad -\half \le \tau_1 \le \half ,
\quad \tau_2 > 0 , \quad |\tau| > 1 \quad ,
\label{eq:metrict}
\end{equation}
with worldsheet coordinates, $\sigma^a$, $a$$=$$1,2$, scaled to
unit length. The combination, ${\sqrt{g}} g^{ab}$, is both diffeomorphism
and Weyl invariant, leading to a gauge-invariant norm for
quadratic deformations of the scalar fields, $X$, as well as for
{\em traceless}
deformations of the metric field, $g_{ab}$. Note that an arbitrary
reparameterization of the metric can be decomposed into trace-dependent and
traceless components, the latter including the effect of a variation in the
worldsheet moduli, $\tau_i$:
\begin{equation}
\delta g_{ab} = g_{ab} \left ( \delta \phi -\nabla_c \delta \sigma^c \right )
-\left( \nabla_a \delta \sigma_b
+ \nabla_b \delta \sigma_a
- g_{ab} \nabla_c \delta \sigma^c \right ) + \delta \tau_i \partial_i g_{ab} \quad .
\label{eq:decompm}
\end{equation}
Thus, the only obstruction to a fully Weyl invariant measure for the string path
integral is the norm for the field $\phi$:
\begin{equation}
|\delta \phi |^2 = \int d^2 \sigma {\sqrt {g}} (\delta \phi)^2  =
\int d^2 \sigma ~ \tau_2 ~ e^{\phi} (\delta \phi)^2 \quad .
\label{eq:phil}
\end{equation}
The natural choice of diffeomorphism invariant norm violates Weyl invariance
explicitly. Fortunately, this obstruction is absent in the critical spacetime
dimension since the worldsheet action turns out to be independent of $\phi$.
The differential operator that maps worldsheet vectors, $\delta \sigma^a$, to
symmetric traceless tensors, usually denoted $(P_1 \delta \sigma)_{ab}$, has two
zero modes, or Killing vectors, on the torus. These are
the constant diffeomorphisms: $\delta \sigma^a_0$. Following
\cite{polyakov}, we invoke the unique diffeomorphism invariant norm on the
tangent space to the space of classical metric configurations at a given
metric, $g_{ab}$. This norm is also
Weyl invariant in the critical dimension \cite{poltorus}:
\begin{equation}
|\delta g_{ab} |^2 = \int d^2 \sigma {\sqrt{g}} \left ( g^{ac} g^{bd} +
C g^{ab} g^{cd} \right ) \delta g_{ab}\delta g_{cd}
, \quad dg dX = J (\tau_i) (d\phi d\delta \sigma)^{\prime} d^2 \tau_i dX \quad ,
\label{eq:normm}
\end{equation}
where the prime denotes exclusion of the zero mode.
The diffeomorphism and Weyl invariant measure for moduli in the string path
integral is derived as shown in \cite{poltorus,dhoker}.
Let us arbitrarily pick the normalization unity for the gaussian path integral
of any field on the worldsheet. This property is assumed to hold for either set
of field variables: $(\delta g,\delta X)$, or
$((\delta \phi\delta\sigma^a )^{\prime} , d^2 \tau_i , \delta X^{\prime})$:
\begin{equation}
1 = \int d\delta g_{ab} e^{-\half |\delta g_{ab}|^2 } =
 \int d\delta \phi e^{-\half |\delta \phi|^2 }=
 \int d\delta \sigma^a e^{-\half |\delta \sigma^a |^2 }=
 \int d\delta X e^{-\half|\delta X |^2 }\quad .
\label{eq:gauss}
\end{equation}
Notice that the arbitrary normalization will drop out in the change of variables,
leaving an unambiguously normalized expression for the Jacobian, $J(\tau_i)$.
In the critical spacetime dimension, $J$ does not depend on $\phi$, and is also
independent of the unknown constant $C$ in Eq.\ (\ref{eq:normm}) \cite{poltorus}.
To see this, we begin by accounting for all of the ordinary gaussian integrals
over constant parameters that contribute to the measure. For the two real worldsheet
moduli, it is easy to check that:
\begin{equation}
1 = (2\pi)^{-1} \left (\int d^2 \sigma {\sqrt{g}} \right )
\int d\tau_1 d \tau_2  e^{-\half (\delta \tau_i)^2 \int d^2 \sigma {\sqrt{g}} }
  \quad ,
\label{eq:gausst}
\end{equation}
which gives the normalization of the integral over moduli.
Likewise, separating out the two real zero modes of the vector Laplacian:
\begin{equation}
d\phi d\delta \sigma^a =
(d\phi d\delta \sigma^a )^{\prime} d \delta \sigma_0^1 d \delta \sigma_0^1
\quad ,
\label{eq:zerom}
\end{equation}
the gaussian integral for the corresponding field variations is
found to be normalized as follows:
\begin{eqnarray}
 1=&& \int d\delta \sigma^1_0 d \delta \sigma_0^2
e^{- \delta \sigma_0^a \delta \sigma^b_0 \int d^2 \sigma {\sqrt{g}}
(\partial_a \phi \partial_b \phi + g_{ab} )}
 \int (d\delta \phi d \delta \sigma^a)^{\prime}
e^{-\half |\delta \phi|^2 -\half |\delta \sigma^a|^2}
\cr
\equiv&& 2\pi \left ( {\rm det} Q_{ab} \right )^{-1/2}
 \int (d\delta \phi d \delta \sigma^a)^{\prime}
e^{-\half |\delta \phi|^2 -\half |\delta \sigma^{a\prime}|^2}
  \quad .
\label{eq:gausstz}
\end{eqnarray}
Finally, we should account for the constant modes of the scalar Laplacian. We
can distinguish the $p+1$ noncompact embedding coordinates
parallel to the Dpbrane worldvolume, assumed to have a common
box regularized volume, from the $25-p$ compact bulk coordinates.
Denoting the size of
each embedding coordinate by $L^{\mu}$, $\mu$ $=$ $0$, $\cdots$,
$25$, we have:
\begin{eqnarray}
1 =&& \int d\delta X  e^{-\half |\delta X|^2 } =
 \prod_{\mu=0}^{25} \int d\delta {\bar{x}}
e^{-\half (\delta {\bar{x}}^{\mu})^2 \int d^2 \sigma {\sqrt{g}} }
 \int d\delta X^{\prime} e^{-\half |\delta X^{\prime} |^2 } \cr
=&&
 (2\pi )^{d/2} \left ( \int d^2 \sigma {\sqrt{g}} \right )^{-d/2}
 \int d\delta X^{\prime} e^{-|\delta X^{\prime} |^2/2 }
 \quad ,
\label{eq:gaussx}
\end{eqnarray}
with $d$ $=$ $26$.
Substituting Eqs.\ (\ref{eq:gausst}), (\ref{eq:gausstz}), and
(\ref{eq:gaussx}), in Eq.\ (\ref{eq:gauss}) gives an unambiguously normalized
expression for the Jacobian $J(\tau_i)$:
\begin{eqnarray}
1 =&& \int d g d X e^{-\half |\delta g|^2 - \half |\delta X|^2 } \cr
=&& \left ( {\rm det}  Q_{ab} \right )^{-1/2}
(2\pi)^{-d/2} \left ( \int d^2 \sigma {\sqrt{g}} \right )^{1+d/2}
\left ({\rm det}^{\prime} {\cal M} \right )^{1/2}  \int
 (d\phi d\delta \sigma)^{\prime} d^2 \tau_i dX^{\prime}
           e^{-\half |\delta g|^2 - \half |\delta X|^2 } ,
\label{eq:jaco}
\end{eqnarray}
where ${\cal M}$ arises from the change of variables, and is a self-adjoint
differential operator on the worldsheet. Its explicit form is obtained by
substitution of Eq.\ (\ref{eq:decompm}) in Eq.\ (\ref{eq:normm}) \cite{poltorus,dhoker}.
We have factored out the redundant integrations over
the gauge parameters $(\delta \phi,\delta \sigma^a)^{\prime}$. Dividing by
the volume of the gauge group to eliminate this redundancy gives the following
simplified expression for the one-loop closed bosonic string vacuum functional
in the critical spacetime dimension:
\begin{eqnarray}
W_{\rm tor}  =&& {{1}\over{{\rm Vol[Diff_0 \times Weyl ]} }}
    \int_{\rm tor} [d g] [d X] e^{ -S[X,g] }
 = \int [d\tau]_{\rm tor} \int [d X ]
   e^{ - S[X, {\hat{g}}] } \cr
 =&&  \prod_{\mu=0}^{25} L^{\mu} \int d^2 \tau
\left ( {\rm det}  Q_{ab} \right )^{-1/2}
(2\pi)^{-d/2} \left ( \int d^2 \sigma {\sqrt{g}} \right )^{1+d/2}
\left ({\rm det}^{\prime} {\cal M} \right )^{1/2}
 \int d\delta X^{\prime} e^{-S[X,{\hat{g}}]} ,
\label{eq:patha}
\end{eqnarray}
where $d$ $=$ $26$.
It remains to perform the integration over embeddings of the closed
worldsurfaces
with toroidal topology, namely, the 26 scalar fields,
$X^{\prime}(\sigma)$. On the torus, the Laplacian on scalars
acts as: $\Delta$$=$$\tau_2^{-2}|\partial_2 $$+$$ \tau \partial_1|^2$.
The orthonormal basis for the scalar field, defined with respect
to the Weyl and diffeomorphism invariant measure \cite{zeta,poltorus},
is given by the complete set of eigenfunctions:
\begin{equation}
X^{\prime} (\sigma^a ) = {\sum_{n_2, n_1}}^{\prime}
             a_{n_2, n_1} \Psi_{n_2 , n_1} (\sigma^a)
, \quad {\rm with} ~
\Psi_{n_2 , n_1 } (\sigma^a) = {{1}\over{{\sqrt{\tau_2}}}}
 e^{2\pi i (n_2 \sigma^2 + n_1 \sigma^1 )} \quad ,
\label{eq:sett}
\end{equation}
and the discrete set of eigenvalues:
\begin{equation}
\omega_{n_2 , n_1 } = 4\pi^2 ( g^{ab} n_a n_b ) =
              {{4\pi^2}\over{\tau_2^2}} | n_2 - \tau n_1 |^2
\quad ,
\label{eq:deegtermt}
\end{equation}
where the subscripts take values in the
range $-\infty$$\le$$n_2$$\le$$\infty$, $-\infty$$\le$$n_1$$\le$$\infty$. Following
Hawking \cite{zeta}, we note that the measure in the tangent space to the
space of embeddings is ultralocal,
a point that has also been stressed by Polchinski \cite{poltorus,dhoker,mine}.
Namely, the functional integral over embeddings, $Z_X (\sigma^a)$, is the
product of ordinary integrals defined at some base point, $\sigma^a$,
on the two-dimensional domain, followed by an integration of the location of the
base point in the domain $0$$\le$$\sigma^a$$\le$$1$. The normalization of the
sum over embeddings, denoted $\mu$ in \cite{zeta}, is determined unambiguously
by the form of the classical action and the gauge invariant norm on the
space of eigenfunctions \cite{poltorus}:
\begin{eqnarray}
&& \int d\delta X^{\prime}
 e^{- {{1}\over{4\pi\alpha^{\prime}}} \int d^2 \sigma X^{\prime} {\sqrt{g}} g^{ab}
\partial_a \partial_b X }
\cr
&& \quad = \mu \prod_{n_2=-\infty}^{\infty}
\prod_{n_1=-\infty}^{\infty \prime }
 \int d a_{n_2 ,  n_1  }
 e^{- \quaft \left ( \sum_{n_2= - \infty}^{\infty }
\sum_{n_1=-\infty }^{\infty \prime} \omega_{n_2, n_1 }
|a_{n_2 ,  n_1 }|^2 \right ) } \cr
&& \quad  = (2 \pi \alpha^{\prime})^{-d/2}
\prod_{n_2=-\infty}^{\infty}
\prod_{n_1=-\infty}^{\infty \prime }
  \left ( \omega_{n_2 , n_1} \right )^{-d/2}
 \equiv (2 \pi \alpha^{\prime})^{-d/2} \left ( {\rm det}^{\prime} \Delta \right )^{-d/2}
\quad .
\label{eq:funcint}
\end{eqnarray}
More significantly, the lack of ambiguity in this normalization is preserved
even after the introduction of a regulator for the infinite products in
Eq.\ (\ref{eq:funcint}). This is due to the fact that the choice of
worldsheet ultraviolet regulator is uniquely determined by the gauge
symmetries.

\vskip 0.1in
One last substitution relates the functional determinants of the vector and scalar
Laplacians \cite{poltorus,dhoker}:
\begin{equation}
\left ( {\rm det}^{\prime} {\cal M} \right )^{1/2} =
\left ({\rm det}^{\prime} ~ 2 ~ \Delta^c_d \right )^{1/2} \left ( {{2}\over{\tau_2^2}} \right )
= {{ 1}\over{2}} {\rm det}^{\prime} \Delta
= \half \prod_{n_2=-\infty}^{\infty}
\prod_{n_1=-\infty}^{\infty \prime } \omega_{n_2 , n_1} \quad .
\label{eq:vect}
\end{equation}
Notice that the dependence of ${\cal M}$ on the unknown constant, $C$, appearing
in Eq.\ (\ref{eq:normm}) drops out in the critical spacetime dimension since the
$\phi$ field decouples \cite{poltorus}.
Substituting Eqs.\ (\ref{eq:funcint}) and (\ref{eq:vect}) in the expression for
the vacuum functional gives the simplified expression:
\begin{eqnarray}
W_{\rm tor}  =
  \prod_{\mu=0}^{25} L^{\mu} \int_F {{d^2 \tau}\over{2\tau^2_2}}
     \tau_2^{-2+1+d/2} (2\pi)^{-d/2} (2\pi \alpha^{\prime} )^{-d/2}
\prod_{n_2=-\infty}^{\infty}
\prod_{n_1=-\infty}^{\infty \prime }
   \left ( \omega_{n_2 , n_1} \right )^{-12}
\quad .
\label{eq:pathx}
\end{eqnarray}

\vskip 0.1in
The infinite product in Eq.\ (\ref{eq:funcint}) can be zeta-function
regulated using a Sommerfeld-Watson integral transform following
\cite{poltorus}. We review this derivation in Appendix C. The result
is \cite{poltorus}:
\begin{equation}
 {\rm det}^{\prime} \Delta = \tau_2^2 e^{-\pi \tau_2/3}
 \prod_{n=1}^{\infty} |1-e^{2\pi i n \tau}|^4
=\tau_2^2 |\eta(q)|^{4}
         \quad .
\label{eq:poldet}
\end{equation}
Substituting in Eq.\ (\ref{eq:pathx}) gives Polchinski's result for
the sum over connected worldsurfaces with the topology of a torus
\cite{poltorus}:
\begin{equation}
W_{\rm tor} =  \prod_{\mu=0}^{25} L^{\mu}
   \int_F {{d^2\tau}\over{2 \tau_2}} (4\pi^2 \alpha^{\prime} \tau_2)^{-13}
        |\eta(\tau)|^{-48} \quad .
\label{eq:poltr}
\end{equation}
It is instructive to compare the zeta-function
regulated expressions in \cite{poltorus} with previous results for the
one-loop closed string amplitudes
obtained in the operator formalism. Friedan's calculation of the closed
string vacuum
amplitude \cite{fried} misses its numerical coefficient. Shapiro's result
for the N-point closed string tachyon scattering amplitude at one loop order
\cite{shap} misses the numerical relation between the one-loop renormalized
closed string coupling and the fundamental string mass scale.
Notice that the appropriate Jacobi theta functions appear in either
worldsheet formalism simply as a consequence of modular invariance. This
symmetry is responsible for the finiteness of one-loop string amplitudes,
and it holds independent of their normalization.

\subsection{One-loop Vacuum Amplitude: Boundaries and Crosscaps}

\vskip 0.1in
There is only one orientable open Riemann surface of Euler number zero, the annulus,
with two boundaries. The corresponding nonorientable surfaces of vanishing Euler
number are
obtained by plugging, respectively, one or both holes of the annulus with a
crosscap. The Mobius strip has a single hole, and the Klein bottle has none.
Thus, the fiducial metric on each nonorientable surface can be chosen identical
to that on the annulus, and the derivation for the gauge-invariant measure for
moduli is unchanged \cite{mine}. The nonorientable surfaces share the same fundamental
domain as the annulus, but {\em the eigenspectrum is appropriately modified by
application of the orientation reversal projection, $\Omega$}. We therefore begin
with a detailed discussion of the annulus. The Mobius strip and Klein bottle will
be obtained as simple modifications of this result.

\vskip 0.1in One further clarification is required to distinguish
results in the presence, or absence, of Dirichlet p-branes. In the
absence of Dpbranes, the boundary of the worldsheet can lie in all
$26$ dimensions, and we impose Neumann boundary conditions on all
$d$$=$$26$ scalars. This gives the traditional open and closed
string theory, whose supersymmetric generalization is the type IB
string theory. T-dualizing $25$$-$$p$ embedding coordinates gives
the open and closed string theory in the background geometry of a
pair of Dpbranes \cite{dbr}. Its supersymmetric generalization is
the type I$^{\prime}$ string theory, when $p$ is even, and the
type IB string theory in generic Dpbrane background, when $p$ is
odd \cite{dbrane}. The Dpbranes define the hypersurfaces bounding
the compact bulk spacetime, which is $(25$$-$$p)$-dimensional.
Since the bulk spacetime has edges, these $25$$-$$p$ embedding
coordinates are Dirichlet worldsheet scalars. It is conventional
to align the Dpbranes so that the distance of nearest separation,
$R$, corresponds to one of the Dirichlet coordinates, call it
$X^{25}$.

\vskip 0.1in In the presence of a pair of Dpbranes, the classical
worldsheet action contributes a background term given by the
Polyakov action for a string of length $R$ stretched between the
Dpbranes \cite{dbrane,polbook,mine}:
\begin{equation}
S_{\rm cl} [G,g] = \quaft \int d^2 \sigma {\sqrt{g}} g^{ab} ~ G^{25,25} (X)
   \partial_a X^{25} \partial_b X^{25}
 =  {{1}\over{2\pi \alpha^{\prime}}} R^2 t \quad ,
\label{eq:backg}
\end{equation}
where the second equality holds in the critical dimension on open world-surfaces
of vanishing Euler number. The background dependence, $e^{-S_{\rm cl}[G,g]}$,
appears in the sum over
connected open Riemann surfaces of any topology, orientable or nonorientable.
Notice that the background action is determined both by the fiducial worldsheet
metric, {\em and} by the bulk spacetime metric, $G_{\mu\nu}[X]$. Notice also that
the boundary of an open worldsheet is now required to lie within the worldvolume
of the Dpbrane, although the worldsheet itself is embedded in all $26$ spacetime
dimensions.

\vskip 0.1in
The metric on the generic annulus can be parameterized by a single real worldsheet
modulus, $t$, and it takes the form:
\begin{equation}
ds^2 = e^{\phi} ( (d\sigma^1)^2 + 4t^2 (d \sigma^2)^2 ), \quad {\sqrt{g}} = e^{\phi} 2t
, \quad 0 \le t \le \infty \quad ,
\label{eq:metrica}
\end{equation}
with worldsheet coordinates, $\sigma^a$, $a$$=$$1,2$, parameterizing a square
domain of unit length. $2t$ is the physical length of either boundary of the
annulus.
The differential operator mapping worldsheet vectors, $\delta \sigma^a$, to
symmetric traceless tensors, usually denoted $(P_1 \delta \sigma)_{ab}$, has
only one zero mode on the annulus. This is
the constant diffeomorphism in the direction tangential to the boundary:
$\delta \sigma^2_0$. Likewise, the analysis of the zero modes of the
scalar Laplacian must take into account the Dpbrane geometry: the $p+1$ noncompact
embedding coordinates satisfying Neumann boundary conditions are treated exactly
as in the case of the torus. The $25-p$ Dirichlet coordinates lack a zero mode.
Thus, the analog of Eq.\ (\ref{eq:gaussx}) reads:
\begin{eqnarray}
1 =&& \int d\delta X  e^{-\half |\delta X|^2 } =
 \prod_{\mu=0}^{p} \int d\delta {\bar{x}}
e^{-\half (\delta {\bar{x}}^{\mu})^2 \int d^2 \sigma {\sqrt{g}} }
 \int d\delta X^{\prime} e^{-\half |\delta X^{\prime} |^2 } \cr
=&&
 (2\pi )^{(p+1)/2} \left ( \int d^2 \sigma {\sqrt{g}} \right )^{-(p+1)/2}
 \int d\delta X^{\prime} e^{-|\delta X^{\prime} |^2/2 }
 \quad .
\label{eq:gaussxa}
\end{eqnarray}
The analysis of the diffeomorphism and Weyl invariant measure for
moduli follows precisely as for the torus, differing only in
the final result for the Jacobian \cite{mine}. The analog of Eq.\ (\ref{eq:jaco})
is given by:
\begin{eqnarray}
1 =&& \int d g d X e^{-\half |\delta g|^2 - \half |\delta X|^2 } \cr
=&& \left ( {\rm det}  Q_{22} \right )^{-1/2}
 {{ \left ( \int d^2 \sigma {\sqrt{g}} \right )^{1/2+(p+1)/2} }\over{
     (2\pi)^{( p+1)/2} }}
\left ({\rm det}^{\prime} {\cal M} \right )^{1/2}  \int
 (d\phi d\delta \sigma)^{\prime} d t  dX^{\prime}
           e^{-\half |\delta g|^2 - \half |\delta X|^2 } ,
\label{eq:jacoa}
\end{eqnarray}
where $({\rm det} Q_{22})$ $=$ $2t$ in the critical dimension, cancelling the
factor of $2t$ arising from the normalization of the integral over the single
real modulus. As shown in \cite{mine}, ${\cal M}$ takes the form:
\begin{equation}
\left ( {\rm det}^{\prime} {\cal M} \right )^{1/2} =
\left ({\rm det}^{\prime} ~ 2 ~ \Delta^c_d \right )^{1/2} \left ( {{1}\over{2t}} \right )
= {{ 1}\over{2}} (2t)^{-1} {\rm det}^{\prime} \Delta
= \half (2t)^{-1} \prod_{n_2=-\infty}^{\infty}
\prod_{n_1=-\infty}^{\infty \prime } \omega_{n_2 , n_1} \quad .
\label{eq:vecta}
\end{equation}
The Laplacian acting on free scalars on an annulus with boundary length
$2t$ takes the form, $\Delta$$=$$(2t)^{-2}\partial_2^2 $$+$$ \partial_1^2$,
with eigenspectrum:
\begin{equation}
\omega_{n_2 , n_1 } = {{\pi^2}\over{t^2}} ( n_2^2 + n_1^2 t^2 )
, \quad \Psi_{n_2,n_1} = {{1}\over{{\sqrt{2t}}}}
 e^{2\pi i n_2 \sigma^2}  {\rm Sin} (\pi n_1 \sigma^1 )
\quad ,
\label{eq:deegtermaf}
\end{equation}
where the subscripts take values in the
range $-\infty$$\le$$n_2$$\le$$\infty$,
and $n_1$$\ge$$0$ for a Neumann scalar, or $n_1$$\ge$$1$ for a
Dirichlet scalar.

\vskip 0.1in In the case of a background electromagnetic field,
${\cal F}_{p-1,p}$, it is convenient to complexify the
corresponding pair of scalars: $Z$$=$$X^p $$+$$i X^{p-1}$
\cite{callt,bachas,dl}. They satisfy the following twisted
boundary conditions:
\begin{eqnarray}
\partial_1 {\rm Re} Z =&& {\rm Im} Z = 0 \quad \sigma^1 = 0
\\
\partial_1 {\rm Re} e^{-i\phi} Z =&& {\rm Im} e^{-i\phi} Z = 0 \quad \sigma^1 = 1
\label{eq:bc}
\end{eqnarray}
Expanding in a
complete set of orthonormal eigenfunctions gives:
\begin{equation}
Z = \sum_{n_2, n_1} z_{n_2, n_1} \Psi_{n_2 , n_1 } = {{1}\over{{\sqrt{2t}}}}
 e^{2\pi i n_2 \sigma^2}  {\rm Sin} \pi (n_1 + \alpha )\sigma^1
\quad ,
\label{eq:set}
\end{equation}
where $\pi\alpha$$=$$\phi$, and $\pi$$-\phi$, respectively
\cite{dl}, and where the subscripts take values in the range
$-\infty$$\le$$n_2$$\le$$\infty$, $n_1$$\ge$$0$. The twisted
complex scalar has a discrete eigenvalue spectrum on the annulus
given by:
\begin{equation}
\omega_{n_2 , n_1 } (\alpha) = {{\pi^2}\over{t^2}} ( n_2^2 + (n_1+\alpha)^2 t^2 )
\quad .
\label{eq:deegterma}
\end{equation}

\vskip 0.1in
Thus, the connected sum over worldsurfaces with the topology of an annulus
embedded in the spacetime geometry of a pair of parallel Dpbranes separated
by a distance $R$ in the direction $X^9$, and in the absence of a magnetic
field, takes the form \cite{mine}:
\begin{equation}
 W_{\rm ann} =  \prod_{\mu = 0}^p L^{\mu}
   \int_0^{\infty} {{dt}\over{2t}}
   (8\pi^2 \alpha^{\prime}t)^{-(p+1)/2} \eta (it )^{-24}
                       e^{-R^2 t /2\pi\alpha^{\prime}}
\quad .
\label{eq:resulttaf}
\end{equation}
In the presence of a worldvolume electromagnetic field, ${\cal
F}_{p-1,p}$, the scalars $X^{p-1}$, $X^{p}$, are complexified.
Substituting the result for the eigenspectrum of the twisted
complex scalar gives:
\begin{equation}
 W_{\rm ann}(\alpha) = \prod_{\mu = 0}^{p-2} L^{\mu}
   \int_0^{\infty} {{dt}\over{2t}}
   (8\pi^2 \alpha^{\prime}t)^{-(p-1)/2} \eta (it )^{-22}
                       e^{-R^2 t /2\pi\alpha^{\prime}}
 {{ e^{\pi t \alpha^2} \eta(it)}\over{ i \Theta_{11} (it \alpha ,it) }}
\quad ,
\label{eq:resultta}
\end{equation}
with $\alpha$$=$$i\phi/\pi$ and $q$$=$$e^{-2\pi t}$.

\vskip 0.1in
The corresponding expressions for the Mobius strip and Klein bottle
are easily derived by making the appropriate orientation projection on
the eigenspectrum on the annulus. For the Mobius strip, and in the
presence of the magnetic field, we have:
\begin{equation}
 W_{\rm mob} = \prod_{\mu=0}^p L^{\mu}
   \int_0^{\infty} {{dt}\over{2t}}
     (16\pi^2 \alpha^{\prime}t)^{-(p-1)/2} \eta (2it )^{-22}
e^{-R^2 t /\pi\alpha^{\prime}} \prod_{n_1=0}^{\infty}
\prod_{n_2=-\infty}^{\infty \prime} \left (  {\rm det} \Delta_{\rm
mob}  \right )^{-1} \quad . \label{eq:rmsannulus}
\end{equation}
The extra factor of two in the contribution from zero modes, and
in the classical action of the stretched string, is easily
understood as follows. Recall that a cylinder is a strip of
boundary length $2t$ and height $1$. The Mobius strip is the strip
twisted by $\Omega$, orientation reversal, and sewn back upon
itself. Thus, the corresponding eigenfunction spectrum is the same
as that on a cylinder {\em with boundary length $4t$}, height $2$,
and twisted boundary conditions on the scalar parameterized by an
angle $\alpha$$=$$\pi/2$. Note that the extra factor of two
cancels out in the integration measure since it has the scale
invariant form $dt/t$. The spectrum of the Laplacian takes the
form:
\begin{eqnarray}
\Psi_{n_2  n_1 } =&& {{1}\over{{\sqrt{2t}}}}
 e^{4\pi i (n_2 + \half) \sigma^2}  {\rm Sin} 2\pi
(n_1 + \half )\sigma^1
\nonumber\\
\Psi_{n_2  n_1 } =&& {{1}\over{{\sqrt{2t}}}}
 e^{4\pi i n_2 \sigma^2}  {\rm Sin} 2\pi (n_1 )\sigma^1
\quad ,
\label{eq:msset}
\end{eqnarray}
where we have separated the $\pi/2$-twisted (odd), and untwisted
(even), eigenfunction sectors on the equivalent annulus.
To obtain the corresponding results in
the presence of the magnetic field, simply introduce the parameter
$\alpha$ in these expressions, where $\alpha$ takes the values,
$\phi/\pi$, or $1$$-$$\phi/\pi$:
\begin{eqnarray}
\Psi_{n_2  n_1 } (\alpha) =&& {{1}\over{{\sqrt{2t}}}}
 e^{4\pi i (n_2 + \half) \sigma^2}  {\rm Sin} 2\pi
(n_1 + \half + \half \alpha )\sigma^1
\nonumber\\
\Psi_{n_2  n_1 } =&& {{1}\over{{\sqrt{2t}}}}
 e^{4\pi i n_2 \sigma^2}  {\rm Sin} 2\pi
(n_1 + \half \alpha )\sigma^1
\quad ,
\label{eq:mssetb}
\end{eqnarray}
The corresponding eigenvalues are:
\begin{eqnarray}
\omega^{\rm odd}_{n_2  n_1 } (\alpha) =&&
 {{4\pi^2}\over{t^2}} (n_2+\half)^2 + 4\pi^2 \left ( n_1+\half+ \half \alpha \right )^2
\nonumber\\
\omega^{\rm even}_{n_2  n_1 } (\alpha) =&&
 {{4\pi^2}\over{t^2}} n_2^2 + 4\pi^2 \left ( n_1 + \half \alpha \right )^2
\quad ,
\label{eq:mseiset}
\end{eqnarray}
where $-\infty$$\le$$n_2$$\le$$\infty$, and $0$$\le$$n_1$$\le$$\infty$.

\vskip 0.1in
Thus, the functional determinant of the complex scalar Laplacian on the Mobius strip
in the presence of the background magnetic field, is given by the infinite product
over both sets of eigenvalues:
gives:
\begin{equation}
{\rm det} \Delta_{\rm mob } =
\prod_{n_2=-\infty}^{\infty} \prod_{n_1=0}^{\infty}
\left [ {{ 4\pi^2}\over{t^2}} n_2^2 + 4\pi^2 \left (n_1 + \half \alpha \right )^2 \right ]
\left [ {{ 4 \pi^2}\over{t^2}} (n_2 + \half) ^2 + 4\pi^2 \left (n_1 + \half + \half \alpha
\right )^2
\right ]
\quad .
\label{eq:mseigenv}
\end{equation}
Regulating the divergent products by the method of zeta-function regularization, as
shown in Appendix C.2, gives the result:
\begin{eqnarray}
{\rm det} \Delta_{\rm mob } =&&
q^{ \half ( \alpha^2 + {{1}\over{6}} -  \alpha) }
 (1-q^{\alpha})^{-1}
 \prod_{n_1=1}^{\infty}
\left [ ( 1- q^{2n_1  - \alpha  })( 1- q^{2n_1 + \alpha  } ) \right ]
\cr
\nonumber \\
\quad &&\quad
\prod_{n_1 =0}^{\infty}
\left [ (1+q^{2n_1 -1 - \alpha })(1+q^{2n_1 -1 + \alpha }) \right ]
\quad .
\label{eq:msresult}
\end{eqnarray}
This can be expressed in terms of the Jacobi theta functions as follows:
\begin{equation}
\left ( {\rm det} \Delta_{\rm mob } \right )^{-1}
 = - i e^{2 \pi t \alpha^2} [{{\eta(2it)}\over{ \Theta_{11} (2it \alpha ,2it) }}]
  [{{\eta(2it)}\over{ \Theta_{00} (2it \alpha ,2it) }}]
\quad ,
\label{eq:msresultt}
\end{equation}
where $\alpha$ denotes $\phi/\pi$.

\vskip 0.1in
Combining with the functional determinants for the free Dirichlet and Neumann
scalars, our final expression for the sum over connected worldsurfaces with the
topology of a Mobius strip is:
\begin{equation}
 W_{\rm mob} = \prod_{\mu=0}^{p}
   \int_0^{\infty} {{dt}\over{2t}}
(16\pi^2 \alpha^{\prime}t)^{-(p+1)/2}
 e^{-R^2 /2\pi\alpha^{\prime}} [\eta(2it) \Theta_{00} (0 ,2it)]^{-12}
\quad .
\label{eq:rmobiusf}
\end{equation}
In the presence of the background magnetic field, the result takes the form:
\begin{eqnarray}
 W_{\rm mob}(\alpha) =&& \prod_{\mu=0}^{p-2}
   \int_0^{\infty} {{dt}\over{2t}}
       (16\pi^2 \alpha^{\prime}t)^{-(p-1)/2}
e^{-R^2 t /2\pi\alpha^{\prime}}
\nonumber\\
\quad && \quad\quad \times
  [\eta(2it) \Theta_{00} (0 ,2it)]^{-10}
  e^{2\pi t \alpha^2} [{{\eta(2it)}\over{ \Theta_{11} (2it \alpha ,2it) }}]
  [{{\eta(2it)}\over{ \Theta_{00} (2it \alpha ,2it) }}]
\quad .
\label{eq:rmobiusm}
\end{eqnarray}

\vskip 0.1in
The amplitude for the Klein bottle follows from similar considerations.
The Klein bottle is
a strip twisted by $\Omega$, orientation reversal, and with {\em both} ends
sewn back upon themselves. Note that the Klein bottle is a closed
worldsurface. Thus, the corresponding eigenfunction spectrum is
the same as that on a cylinder with boundary
length $4t$, height $2$, but with periodicity imposed on {\em both} edges.
The periodicity condition implies that we include both left- and
right-moving annulus modes, and with equal weight. For convenience,
we present the results directly in the presence of the background
magnetic field:
\begin{equation}
\Psi_{n_2  n_1 } (\alpha) = {{1}\over{{\sqrt{2t}}}}
 e^{4\pi i n_2 \sigma^2}  {\rm Sin} 2\pi
(n_1 + \half \alpha )\sigma^1
\quad ,
\label{eq:setkb}
\end{equation}
and the eigenvalues are given by:
\begin{equation}
\omega_{n_2  n_1 } =
 {{4\pi^2}\over{t^2}} n_2^2 + 4\pi^2(n_1+ \half \alpha)^2
\quad ,
\label{eq:eiset}
\end{equation}
with the usual range for $n_1$ and $n_2$.
The functional determinant takes the form:
\begin{equation}
{\rm det} \Delta_{\rm kb } =
\prod_{n_2=-\infty}^{\infty} \prod_{n_1=0}^{\infty}
\left [ {{ 4\pi^2}\over{t^2}} n_2^2 + 4\pi^2 (n_1 + \half \alpha )^2 \right ]
\quad .
\label{eq:eigenv}
\end{equation}
Following zeta-function regularization
of the divergent eigenvalue sums, the result for a
complex twisted scalar takes the form:
\begin{eqnarray}
\left ( {\rm det} \Delta_{\rm kb } \right )^{-1}
=&& \left ( q^{  \alpha^2 + {{1}\over{6}} -  \alpha }
 (1-q^{2\alpha})^{-1}
 \prod_{n_1=1}^{\infty}
\left [ ( 1- q^{2n_1  - \alpha  })( 1- q^{2n_1 + \alpha  } ) \right ] \right )^{-1}
\cr
\nonumber\\
 =&& e^{2\pi t \alpha^2} [{{\eta(2it)}\over{ \Theta_{11} (2it \alpha ,2it) }}]
\quad .
\label{eq:reskb}
\end{eqnarray}
Combining with the contributions from the free Dirichlet and Neumann
scalars, we obtain:
\begin{equation}
 W_{\rm kb} =  \prod_{\mu=0}^{p}
L^{\mu} \int_0^{\infty} {{dt}\over{2t}} (16 \pi^2
\alpha^{\prime}t)^{-(p+1)/2} \eta (2it )^{-24} e^{-R^2 t
/\pi\alpha^{\prime}} \quad . \label{eq:rkbs}
\end{equation}
In the presence of the background magnetic field, the corresponding result is:
\begin{equation}
 W_{\rm kb} (\alpha) = \prod_{\mu=0}^{p-2} L^{\mu}
   \int_0^{\infty} {{dt}\over{2t}} (16\pi^2 \alpha^{\prime}t)^{-(p-1)/2} \eta (2it )^{-22}
e^{-R^2 t /\pi\alpha^{\prime}}
  e^{2\pi t \alpha^2} [{{\eta(2it)}\over{ \Theta_{11} (2it \alpha ,2it) }}]
\quad .
\label{eq:rkbsb}
\end{equation}

\section{Type I-I$^{\prime}$ String Theory in an Electromagnetic Background}

The derivation of the gauge invariant measure for moduli given in
Appendix A can be easily extended to the case of the
supersymmetric unoriented open and closed string theories, type I
and type I$^{\prime}$ string theory \cite{dhoker,polbook}. We
begin with the contribution from worldsurfaces with the topology
of an annulus in the presence of a background electromagnetic
field, and in the background spacetime geometry of a pair of
Dpbranes separated by a distance $R$. The first principles
derivation of the result for the annulus was derived in
\cite{wils}. We will also derive the results for the sum over
unoriented world-surfaces with the topology of a Mobius strip, and
a Klein bottle, in the discussion that follows below. Beginning
with the annulus:
\begin{eqnarray}
 W_{ann-I} (\alpha) =&&  \prod_{\mu=0}^{p-2}
   \int_0^{\infty} {{dt}\over{2t}} (8\pi^2 \alpha^{\prime}t)^{-(p-1)/2}
\eta (it )^{-6} e^{-R^2 t /2\pi\alpha^{\prime}}
  {{ e^{\pi t \alpha^2} \eta(it)}\over{ i \Theta_{11} (it \alpha ,it) }}
\nonumber\\
\quad && \quad \times
\prod_{n_2=0}^{\infty} \prod_{n_1=-\infty}^{\infty \prime}
\left (  {\rm det}_{\rm ann-I} \Delta_{n_2 + \half , n_1 + \half + \alpha } \right )^{1}
\left (  {\rm det}_{\rm ann-I} \Delta_{n_2 + \half , n_1 + \half } \right )^{3} ,
\label{eq:rannulusf}
\end{eqnarray}
where we have included the contribution from worldsheet bosonic fields derived in
the previous section.

\vskip 0.1in
Let us understand the eigenvalue spectrum of the worldsheet fermions in more detail.
Recall that the functional determinant of the two-dimensional
Dirac operator acting
on a pair of Majorana Weyl fermions satisfying twisted boundary conditions
is equivalent, by Bose-Fermi equivalence, to the functional determinant
of the scalar Laplacian {\em raised to the inverse power}. This provides
the correct statistics. In addition, we have the constraint of world-sheet
supersymmetry. This requires that the four complexified Weyl fermions satisfy
identical boundary conditions in each sector of the theory in the $\sigma^1$
direction. For a complex Weyl
fermion satisfying the boundary condition:
\begin{eqnarray}
\psi (1,\sigma^2) =&& - e^{\pi i a} \psi(0,\sigma^2)
\nonumber\\
\psi (\sigma^1 , 1 ) =&& - e^{ \pi i b} \psi(\sigma^1 , 0)
\quad ,
\label{eq:ferbc}
\end{eqnarray}
the Bose-Fermi equivalent scalar eigenspace takes the form:
\begin{equation}
\Psi_{n_2 + \half , n_1 + \half (1+a)  } = {{1}\over{\sqrt{2t}}} e^{ 2\pi i (n^2 + \half
(1\pm b) )\sigma^2}
{\rm Cos} \pi (n_1 + \half (1 \pm a) ) \quad ,
\label{eq:fermseig}
\end{equation}
where we sum over $-\infty$$\le$$n_2$$\le$$\infty$, $n_1$$\ge$$0$.
Notice that the unrotated oscillators are, respectively, half-integer
or integer moded as expected for the scalar equivalent of antiperiodic
or periodic worldsheet fermions.
Finally, we must sum over periodic and antiperiodic sectors, namely,
with $a$, $b$ equal to $0$, $1$.
As reviewed in the appendix, weighting the $(a , b)$ sector
of the path integral by the factor $e^{  \pi i a b}$
gives the following result for the fermionic partition function
\cite{polbook}:
\begin{eqnarray}
Z^{a}_{b} (\alpha , q) =&&
q^{\half a^2 - {{1}\over{24}}} e^{  \pi i ab}
 \prod_{m=1}^{\infty} \left [ ( 1+ e^{  \pi i b } q^{m - \half(1 - a) + \alpha })
  ( 1+ e^{- \pi i b } q^{m - \half (1+ a)  - \alpha } )\right ]
\nonumber\\
\quad \equiv&&  {{1}\over{e^{\pi t \alpha^2} \eta(it) }} \Theta_{ a    b }  (\alpha it , it )
\quad .
\label{eq:theta}
\end{eqnarray}
We have included a possible rotation by $\alpha$ or $1$$-$$\alpha$ as in the
previous section. This applies for the Weyl fermion partnering the twisted complex
worldsheet scalar. Substituting in the path integral, and summing over
$a$, $b$$=$$0$, $1$,
for all fermions,
and over $\alpha$ and $1$$-$$\alpha$ for the Weyl fermion partnering
the twisted complex scalar, gives the result:
\begin{eqnarray}
 W_{\rm ann-I} &&= \prod_{\mu=0}^{p-2} L^{\mu}
   \int_0^{\infty} {{dt}\over{2t}}
(8\pi^2 \alpha^{\prime}t)^{-(p-1)/2}
      e^{-R^2 t /2\pi\alpha^{\prime}} \times \left \{ \eta (it )^{-6}
 \left [ {{ e^{\pi t \alpha^2} \eta(it)}\over{ \Theta_{11} (it \alpha ,it) }}
\right ] \right \}
\nonumber\\
\quad && \times
\left [ {{ \Theta_{0 0 } (it \alpha , it)}\over{ e^{\pi t \alpha^2} \eta (it) }}
 \left ( {{ \Theta_{00} (0 , it)}\over{ \eta (it) }} \right )^3
-
 {{ \Theta_{0 1 } (it\alpha , it)}\over{ e^{\pi t \alpha^2} \eta (it) }}
 \left ( {{ \Theta_{0 1} (0 , it)}\over{ \eta (it) }} \right )^3
-
 {{ \Theta_{1 0} (it\alpha , it)}\over{ e^{\pi t \alpha^2} \eta (it) }}
 \left ( {{ \Theta_{10} (0 , it)}\over{ \eta (it) }} \right )^3
\right ]
\nonumber \\
\label{eq:rannulusfer}
\end{eqnarray}
where we have used the fact that $\Theta_{11 }(0,it)$ equals zero.
The factor within curly brackets is the contribution from
world-sheet bosons; that within square brackets in the second line
of the expression is the contribution from world-sheet fermions.

\vskip 0.1in
Likewise, we can write down the corresponding results for the sum over worldsurfaces
with the topology of Mobius strip or Klein bottle by invoking Bose-Fermi equivalence,
and by using the appropriate twisted cylinder eigenspaces. The detailed derivation is
outlined in appendix C. For the Mobius strip we have the result:
\begin{eqnarray}
 W_{\rm mob-I} =&& \prod_{\mu=0}^{p-2}
   \int_0^{\infty} {{dt}\over{2t}}
(16\pi^2 \alpha^{\prime}t)^{-(p-1)/2}
 e^{-R^2 t /\pi\alpha^{\prime}} \times \left \{
 [\eta(2it)]^{-6}[{{e^{2\pi t \alpha^2 } \eta(2it)}\over{ \Theta_{11} (2it \alpha ,2it)
  }}] \right \}
\nonumber\\
\quad &&\times
   \left \{[{{\eta(2it)}\over{\Theta_{00} (0 ,2it)}}]^{3}
  [{{e^{2\pi t \alpha^2} \eta(2it)}\over{ \Theta_{00} (2it \alpha ,2it)
  }}]\right \} \times \left \{
\left [ {{ \Theta_{01 } (2it \alpha , 2it)}\over{ e^{2\pi t
\alpha^2} \eta (2it) }} {{ \Theta_{10 } (2 it \alpha , 2it)}\over{
e^{2\pi t \alpha^2} \eta (2it) }}\right ] \right \}
\nonumber\\
\quad &&\quad\quad \times
   \left [ \left ( {{ \Theta_{01} (0 , 2it)}\over{ \eta (2it) }}
\right )^3
 \left ( {{ \Theta_{10} (0 , 2it)}\over{ \eta (2it) }} \right )^3
~-~  \left ( {{ \Theta_{10} (0 , 2it)}\over{ \eta (2it) }} \right )^3
 \left ( {{ \Theta_{01} (0 , 2it)}\over{ \eta (2it) }} \right )^3   \right
 ]
\quad .
\label{eq:rmobiusfs}
\end{eqnarray}
We have used the fact that $\Theta_{11}(0,2it)$ vanishes. The
factor within curly brackets in the first line of this expression
is the contribution from the eight transverse bosonic modes, in
the presence of the electromagnetic background. The second, and
third, lines of the expression are the contributions from the
worldsheet fermions. Notice that the Ramond-Ramond, and
Neveu-Schwarz-Neveu-Schwarz, sectors give identical contributions
of opposite sign.

\vskip 0.1in
The corresponding result for the Klein bottle is:
\begin{eqnarray}
 W_{\rm kb-I} =&& \prod_{\mu=0}^{p-2} L^{\mu}
   \int_0^{\infty} {{dt}\over{2t}}
(16\pi^2 \alpha^{\prime}t)^{-(p-1)/2}
 e^{-R^2 t /\pi\alpha^{\prime}} \times \left \{ [\eta (2it )]^{-6}
   [{{e^{2\pi t \alpha^2} \eta(2it)}\over{ \Theta_{11} (it \alpha ,2it)
   }}]\right \}
\nonumber\\
\quad && \quad \times
 [
 \left ({{\Theta_{00} (0 , 2it)}\over{\eta(2it)}}\right )^3
 {{\Theta_{00} ( 2it \alpha , 2it)}\over{e^{2\pi t \alpha^2} \eta(2it)}}
- \left ({{ \Theta_{10}(0,2it)} \over{\eta(it)}} \right )^3
  {{\Theta_{10} ( 2it \alpha , 2it)}\over{e^{2\pi t \alpha^2} \eta(2it)}}
\nonumber\\
\quad &&\quad\quad\quad
- \left ({{ \Theta_{01}(0,2it)} \over{\eta(it)}} \right )^3
  {{\Theta_{01} ( 2it \alpha , 2it)}\over{e^{2\pi t \alpha^2} \eta(2it)}}
 ] .
\label{eq:rkbsf}
\end{eqnarray}
For simplicity, setting $\alpha$$=$$0$, let us write down the
result for the sum over connected one-loop vacuum graphs in the
Dpbrane background geometry, but without an electromagnetic field.
The sum over worldsurfaces with the topology of a torus decouples
from the sum over worldsurfaces with boundary and/or crosscap,
since it is insensitive to the Dpbranes. It also vanishes as a
consequence of the unbroken spacetime supersymmetry. The sum over
unoriented connected worldsurfaces with vanishing Euler character
in type I(I$^{\prime}$) string theory in the background of a pair
of parallel and static Dpbrane stacks, each with $N$ coincident
Dpbranes, takes the form:
\begin{eqnarray}
 W_{\rm I} &&=  \prod_{\mu=0}^{p} L^{\mu}
   \int_0^{\infty} {{dt}\over{2t}}
(8\pi^2 \alpha^{\prime}t)^{-(p+1)/2} e^{-R^2 t
/2\pi\alpha^{\prime}}
\nonumber\\
\quad && \quad \times \{
 {{N^2 }\over{\eta(it)^8 }} \left [
\left ({{ \Theta_{00} (0 , it)}\over{ \eta (it) }} \right )^4
- \left ( {{ \Theta_{0 1} (0 , it)}\over{ \eta (it) }} \right )^4
- \left ( {{ \Theta_{10} (0 , it)}\over{ \eta (it) }} \right )^4 \right ]
\nonumber\\
&& \quad
  - {{2^{6} N}\over{\eta(it)^8}} \left ( {{\eta (it)}\over{\Theta_{00}(it) }}\right )^{4}
\left [ \left ( {{\Theta_{01} (0,it)}\over{ \eta (it) }} \right
)^4
 \left ( {{\Theta_{10} (0,it)}\over{ \eta (it ) }}\right )^{4}
- \left ( {{\Theta_{10} (0,it)}\over{ \eta (it) }}\right )^{4}
 \left ( {{\Theta_{01} (0,it)}\over{ \eta (it) }}\right )^{4} \right ]
\nonumber\\
&& \quad
   +  {{2^{10}}\over{\eta(it)^8}} \left [
 \left ({{\Theta_{00} (0 , it)}\over{\eta(it)}}\right )^4
- \left ({{ \Theta_{10}(0,it)} \over{\eta(it)}} \right )^4 - \left
({{ \Theta_{01}(0,it)} \over{\eta(it)}} \right )^4 \right ]  \}
\quad , \label{eq:RRcharge}
\end{eqnarray}
where we have used the fact that $t_{\rm MS}$$=$$2t_{\rm ann}$,
$t_{\rm KB}$$=$$2t_{\rm ann}$. $\Theta_{11}(0,it)$ vanishes as a
consequence of the zero mode in the Ramond-Ramond sector for
worldsheet fermions \cite{polbook}. $W_{\rm I} $ vanishes as a
consequence of target spacetime supersymmetry, as can be seen by
use of the abstruse identity relating the Jacobi theta functions
in the zero external field annulus, and Klein bottle, amplitudes
\cite{polbook}; notice that the Mobius strip gives a vanishing
contribution even in the presence of the external field. The
generalization of Eq.\ (\ref{eq:RRcharge}) in the presence of an
external electromagnetic field is obtained by combining the
expressions in Eqs.\ (\ref{eq:rannulusfer}), (\ref{eq:rmobiusfs}),
and (\ref{eq:rkbsf}).

\vskip 0.1in It is important to understand that in the type I
vacuum with $N$$=$$32$ Dpbranes, the vanishing of the one-loop
vacuum amplitude in zero external field can also be understood as
a consequence of reasons having nothing whatsoever to do with
target spacetime supersymmetry.\footnote{This is the basic
observation used by us in recent works \cite{holo,micro} in
arriving at an expression for the free energy of the canonical
ensemble of type I strings: target spacetime supersymmetry is
broken by the introduction of thermal phases in the finite
temperature vacuum. But the one-loop string free energy
nevertheless vanishes as a consequence of R-R sector tadpole
cancellation alone.} As has been stressed by Polchinski
\cite{openbs,dbrane,dnotes,polbook}, the number of Dpbranes is
fixed to be $32$ by the requirement of Ramond-Ramond sector
tadpole cancellation: $N-2^5$$=$$0$, ensuring the absence of a
propagating unphysical Ramond-Ramond state. The latter fact was
first noted in \cite{openbs}. In order to see this, note that the
factor within curly brackets in lines 2--4 of the expression in
Eq.\ (\ref{eq:RRcharge}) is the partition function of the
worldsheet superconformal field theory (SCFT) on the strip of
length $2t$, with appropriate orientation projection on the
spectrum in the case of the Mobius strip and Klein bottle
contributions, as explained above. Expanding in powers of
$e^{-2\pi t}$ yields the so-called $q$--expansion of the partition
function of the boundary SCFT. The leading terms of order $e^{\pi
t}$ vanish, a precise cancellation indicating the absence of
target spacetime tachyons in the open string mass spectrum. The
next order, $q^0$, counts the number of massless target spacetime
bosons and target spacetime fermions, contributing with opposite
signs to the one-loop string vacuum functional. If we collect the
contributions at this order from target spacetime bosons alone,
namely, the $O(q^0)$ term in the NS-NS sector of the annulus,
Mobius strip, and Klein bottle amplitudes, we find a precise
cancellation because the overall numerical coefficient vanishes
when $N$$=$$2^5$ \cite{openbs,dnotes,polbook}:
\begin{equation}
 W_{\rm I (NS-NS)} = \half (N-2^5)^2 \left [ \prod_{\mu=0}^{p} L^{\mu} \right ]
(8\pi^2 \alpha^{\prime})^{-(p+1)/2} \int_0^{\infty} dt
t^{-(p+1)/2} e^{-R^2 t /2\pi\alpha^{\prime}} \left [ 16 + O(e^{\pi
t}) \right ] \quad , \label{eq:lead}
\end{equation}
and where $W_{\rm I(R-R)}$ is an expression that takes identical
form, but with opposite overall sign \cite{openbs,dnotes,polbook}.
Finally, we should note that the contribution from the annulus
amplitude to $W_{\rm I}$ from the NS-NS sector alone, namely, from
just target spacetime bosons, carries the significant information
about the charge of the Dirichlet-pbrane \cite{dbrane}. This is
explained at length in section 2.3.

\section{\bf Zeta-function Regularization of Infinite Sums}

The regularization of a divergent sum over the discrete
eigenvalue spectrum of a self-adjoint differential operator by the
zeta function method can always be carried out in closed form when the
eigenvalues are known explicitly \cite{zeta}.
This is the case for all of the infinite sums encountered in one-loop
string amplitudes \cite{poltorus,polbook,mine}.
We illustrate the basic method with a review of Polchinski's
calculation of the zeta-regularized functional determinant of the
scalar Laplacian on the torus \cite{poltorus}:
\begin{equation}
{\rm ln} ~ {\rm det}^{\prime} \Delta_{\rm tor} =
\lim_{m\to 0}
\sum_{n_1 =-\infty}^{\infty} \sum_{n_2 = -\infty}^{\infty}
  {\rm ln} \left [ {{4\pi^2}\over{\tau^2}}
( n_2 - \tau n_1  )
(n_2 - {\bar{\tau}} n_1 ) + m^2 \right ]
 -  {\rm ln} \left [ {{4\pi^2 m^2}\over{\tau_2^2}} \right ]
\quad .
\label{eq:nfprot}
\end{equation}
The $n_1$$=$$n_2$$=$$0$ term has been included in the infinite sum
by introducing an infrared regulator mass, $m$, for the zero mode.
We will take the limit $m$$\to$$0$ at the end of the calculation.
Following Hawking \cite{zeta}, we begin by expressing the first term in
Eq.\ (\ref{eq:nfprot}) in the equivalent form:
\begin{equation}
S_{\rm tor} = - \lim_{s,m \to 0} {{d}\over{ds}}
\left \{ ({{4\pi^2}\over{\tau_2^2 }})^{-s}
\sum_{n_2,n_1=-\infty}^{\infty }
\left [ ( n_2 - \tau n_1  )
(n_2 - {\bar{\tau}} n_1 )
+ m^2 \right ]^{-s} \right \}
\quad .
\label{eq:nfprost}
\end{equation}
Notice that the infinite sums are
manifestly convergent for ${\rm Re}$ $s$ $>$ $1$. The
required $s$ $ \to$ $0$ limit can be obtained by analytic continuation
in the variable $s$. The analogous step for the second term in Eq.\
(\ref{eq:nfprot}) yields the relation:
\begin{equation}
 + \lim_{m \to 0} \lim_{s \to 0} {{d}\over{ds}}
[{{4\pi^2 m^2  }\over{\tau_2^2 }}]^{-s}
=  2~ {\rm log} ~ \tau_2 - \lim_{m \to 0}  2 ~ {\rm log} ~ (2\pi m )
 \quad .
\label{eq:zerot}
\end{equation}
The finite term in this expression contributes the overall factor of $\tau_2^2$
to the result given in Eq.\ (\ref{eq:poldet}).

\vskip 0.1in
The infinite summation over $n_2$ is carried out using a Sommerfeld-Watson
transform as in \cite{poltorus}. We invoke the Residue Theorem in giving
the following contour integral representation of the infinite sum as
follows:
\begin{equation}
\sum_{n = -\infty }^{\infty} \left [ n^{2} + x^2 \right ]^{-s} =
\oint_{\sum_n {\cal C}_n} {{dz}\over{2\pi i}} \pi {\rm cot} (\pi
z)
  \left ( z^2 + x^2 \right )^{-s}
\quad ,
\label{eq:contoureigt}
\end{equation}
where ${\cal C}_n$ is a small circle enclosing the pole at $z$$=$$n$
in the counterclockwise sense. The contours may be deformed without encountering any
new singularities into the pair of straight line contours, ${\cal C}_{\pm}$,
where the line ${\cal C}_+$ runs from $\infty $$+$$i \epsilon$ to
$-\infty$$+$$i\epsilon$, connecting smoothly to the line ${\cal C}_-$, which runs from
$-\infty$$-$$i\epsilon$ to $\infty$$-$$i\epsilon$.

\vskip 0.1in
Alternatively, we can choose to close the contours, ${\cal C}_{\pm}$, respectively,
in the upper, or lower, half-planes along the circle of infinite radius. Note that
the integrand has
additional isolated poles in the complex plane at the points $z$$=$$\pm$$ix$.
We will make the following substitution in the integrand:
\begin{eqnarray}
{{{\rm Cot}(\pi z)} \over {i}}=&& {{2e^{ i\pi z}}\over{e^{i\pi z}-e^{-i \pi z}}}
        - 1 \quad , \quad {\rm Im }~z > 0
\nonumber \\
{{{\rm Cot}(\pi z)} \over {i}}=&& {{-2e^{- i\pi z}}\over{e^{i\pi z}-e^{-i \pi z}}}
        + 1  \quad , \quad {\rm Im }~z < 0
\label{eq:cotid}
\end{eqnarray}
when the contour is to be closed, respectively, in the upper, or lower, half-plane.
This ensures that the integrand is convergent at all points within the enclosed region
other than the isolated poles. Thus, we obtain the following alternative contour
integral representation of the infinite sum over $n_2$,
setting $x^2$$=$$n_1^2 \tau_2^2 $$+$$ m^2$:
\begin{eqnarray}
\sum_{n_2 = -\infty}^{\infty}
\left [ (n_2 - n_1 \tau_1)^{2} + x^2 \right ]^{-s}
=&& \oint_{{\cal C}_+} dz
 \left [ {{e^{ i\pi z}}\over{e^{i\pi z}-e^{-i \pi z}}} - \half \right ]
\left ( (z-n_1 \tau_1)^2 + x^2 \right )^{-s}
\nonumber \\
&&\quad + \oint_{{\cal C}_-} dz
 \left [ -{{e^{ - i\pi z}}\over{e^{i\pi z}-e^{-i \pi z}}} + \half \right ]
\left [ (z - n_1\tau_1)^2 + x^2 \right ]^{-s}
\quad .
\label{eq:contoureiget}
\end{eqnarray}
Note that the contours are required to avoid the
branch cuts which run, respectively, from
$+$$ix$ to $+$$i\infty$, and from
$-$$ix$ to $-$$i\infty$.
Let us evaluate these integrals as before.
The constant pieces from the square brackets in Eq.\ (\ref{eq:contoureiget})
combine to give:
\begin{equation}
I_1 (s , x) = \half \left ( \int_{{\cal C}_- } dz -
 \int_{{\cal C}_+} dz \right )  ((z-n_1\tau_1)^2 + x^2 )^{-s}
 = x^{-2s+1} \int_{-\infty}^{\infty} du (1+ u^2 )^{-s}
\quad .
\label{eq:finttranseigf}
\end{equation}
The integral simply yields the beta function, $B(\half, s-\half)$.
Taking the $s$-derivative followed by the $s$$=$$0$ limit gives,
\begin{equation}
\lim_{s\to 0} {{d}\over{ds}}
 x^{-2s+1} B(\half , s- \half )
= \lim_{s\to 0} {{d}\over{ds}}
\half x^{-2s+1} {{{\rm sin} (\pi s)}\over{{\sqrt{\pi}}}}
\Gamma (1-s) \Gamma (s - \half )
= - 2 \pi x
\quad .
\label{eq:finttranseigk}
\end{equation}
Substituting for $x^2$$=$$n^2_1 \tau_2^2$$+$$m^2$,
and taking the $m$$=$$0$ limit, gives
\begin{equation}
-4\pi \tau_2 \sum_{n_1=1}^{\infty} n_1 ~-~ 2 \pi ~ \lim_{m\to 0} m
~=~ - 4 \pi \tau_2 \zeta (-1) =  2 \pi \tau_2 B_2(0) = {{1}\over{3}} \pi \tau_2
\quad ,
\label{eq:tott}
\end{equation}
where we have used $B_2(q)$$=$$q^2$$-$$q$$+$${{1}\over{6}}$.

\vskip 0.1in
Next, we tackle the non-constant pieces from the square brackets in
Eq.\ (\ref{eq:contoureiget}). It is helpful to take the $s$-derivative and
the $s$ $\to$ $0$ limit prior to performing the contour integral.
We begin with the contour integral in the upper half-plane:
\begin{eqnarray}
I_2 ( x, s) =&&
\int_{{\cal C}_+} dz
 {{e^{ i\pi z}}\over{e^{i\pi z}-e^{-i \pi z}}}
  \left ( (z-n_1 \tau_1)^2 + x^2 \right )^{-s}
\nonumber \\
        =&&
- 2 {\rm sin} (\pi s) \int_{x }^{\infty} dy
 {{e^{ -\pi y  }}\over{e^{\pi y } -e^{- \pi y } }}
  \left ( (y+in_1\tau_1)^2 - x^2 \right )^{-s}
         \quad .
\label{eq:twfsumtrans2eigt}
\end{eqnarray}
Taking the derivative with respect to $s$, and setting $s$$=$$0$,
gives:
\begin{eqnarray}
{{d}\over{ds}} I_2 ( x, s)|_{s=0} =&&
- 2 \pi \int_{ x-in_1\tau_1 }^{\infty} dy
 {{e^{ -\pi y  }}\over{e^{\pi y } -e^{- \pi y } }}
\nonumber \\
=&& - 2  {\rm log} \left | 1 - e^{-2\pi (x-in_1\tau_1)} \right |
\quad .
\label{eq:twequaeigt}
\end{eqnarray}
The ${\cal C}_-$ integral gives an identical contribution since,
\begin{eqnarray}
I_3 ( x, s) =&&
- \int_{{\cal C}_-} dz
 {{e^{ -i\pi z}}\over{e^{i\pi z}-e^{-i \pi z}}}
  \left ( (z-n_1\tau_1)^2 + x^2 \right )^{-s}
\nonumber \\
        =&&
 - 2 {\rm sin} \pi s \int_{ x }^{\infty} dy
 {{e^{ -\pi y  }}\over{e^{\pi y} -e^{- \pi y } }}
  \left ( (y+in_1\tau_1)^2 - x^2 \right )^{-s}
         \quad .
\label{eq:fsumtrans2eigt}
\end{eqnarray}
Upon taking the $s$-derivative, and setting $s$$=$$0$,
we get the same result as in Eq.\ (\ref{eq:twfsumtrans2eigt}):
\begin{eqnarray}
{{d}\over{ds}} I_3 ( x, s)|_{s=0} =&&
- 2 \pi \int_{ x -in_1\tau_1}^{\infty} dz
 {{e^{ -\pi y }}\over{e^{\pi y } -e^{- \pi y } }}
\nonumber \\
=&& - 2 {\rm log} \left | 1 - e^{-2\pi (x-in_1\tau_1)} \right |
\quad .
\label{eq:equaeig}
\end{eqnarray}
Combining all of the contributions to $S_{\rm tor}$
gives the following result in the $m$$\to$$0$ limit:
\begin{equation}
S_{\rm tor} = -  {{\pi \tau_2  }\over{3}}
+ 4 \sum_{n_1=1}^{\infty} {\rm log} |1- e^{2 \pi i n_1 \tau }|
- \lim_{m\to 0} \left [  \pi m  ~+~ 2~ {\rm log} (2\pi m) ~-~
 2 {\rm log} (1- e^{-2 \pi m } ) \right ]
\quad ,
\label{eq:equsumt}
\end{equation}
Notice that the divergent terms in the $m$$\to$$0$ limit cancel
as is seen by Taylor expanding the logarithm in the last term.

\vskip 0.1in
Combining with the result from Eq.\ (\ref{eq:zerot}) gives Polchinski's
result for the functional determinant of the scalar
Laplacian on the torus:
\begin{equation}
\prod_{n_2=-\infty}^{\infty } \prod_{n_1 = -\infty}^{\infty \prime}
\omega_{n_1 n_2}  =
 \tau_2 e^{-\pi \tau_2/3}
\prod_{n_1=1}^{\infty} | 1- q^{n_1 } |^{4}
\quad ,
\label{eq:resultt}
\end{equation}
where $q$$=$$e^{2\pi i \tau }$.

\subsection{Annulus: Twisted Complex Scalar Eigenspectrum}

The eigenspectrum of the scalar Laplacian on a surface with
boundary includes a dependence on an electromagnetic background,
reflected as a twist in the boundary conditions satisfied by the
scalar. As an illustration, let us work out the functional
determinant of the scalar Laplacian for worldsheets with the
topology of an annulus. The case of the Mobius strip and Klein
bottle are simple extensions which do not introduce any
significant new feature into the nature of the infinite summation.
Since the required sums only differ in the choice of \lq\lq
twist", the results can be straightforwardly written down given
the result for the annulus with generic twist $\alpha$.

\vskip 0.1in
Begin with the eigenspectrum on the annulus. In the case of the free
Neumann scalars, we must introduce an infrared regulator mass for the
zero mode, as shown in the case of the torus.
The functional determinant of the Laplacian can be written in the form
\cite{mine}:
\begin{equation}
{\rm ln} ~ {\rm det}^{\prime} \Delta =
\lim_{m \to 0}
  \sum_{n_1 =0}^{\infty} \sum_{n_2 = -\infty }^{\infty}
  {\rm log} \left [ {{\pi^2}\over{t^2}} ( n_2^2 + n_1^2 t^2 + m^2  ) \right ]
 -  {\rm log} \left ( {{\pi^2 m^2}\over{t^2}} \right )
\quad .
\label{eq:nfproaa}
\end{equation}
The first term in Eq.\ (\ref{eq:nfproaa}) is a special case of the infinite
sum with generic twist, $\alpha$, and the zeta-regulated result can be obtained
by setting $\alpha$$=$$0$ in the generic calculation which will be derived below.
The second term in Eq.\ (\ref{eq:nfproaa}) yields the result:
\begin{equation}
 + \lim_{m \to 0} \lim_{s \to 0} {{d}\over{ds}}
[{{4\pi^2 m^2  }\over{4t^2 }}]^{-s}
=  2~ {\rm log} ~ 2t
 - \lim_{m \to 0}
   - 2 ~ {\rm log} ~ (2\pi m )
 \quad .
\label{eq:zerota}
\end{equation}
This contributes the correct power of $2t$ to the measure of the path integral
for a free Neumann scalar \cite{mine}. For the Dirichlet scalar, we must remember
to drop the $n_1$$=$$0$ modes from the double sum above since the sine eigenfunction
vanishes for all values of $\sigma^1$, not only at the boundary. Thus, the $n_1$
summation begins from $n_1$$=$$1$.

\vskip 0.1in
Now consider the case of the twisted Neumann scalar. There is no need to introduce
an infrared regulator in the presence of a magnetic field since there
are no zero modes in the eigenvalue spectrum. Thus, the analysis of the infinite
eigensum is similar to that for a Dirichlet scalar, other than the incorporation of
twist. We begin with:
\begin{equation}
S_{\rm ann} = - \lim_{s \to 0} {{d}\over{ds}}
[{{\pi^2}\over{t^2 }}]^{-s}
\sum_{n_2=-\infty}^{\infty\prime }
\sum_{n_1 = 1}^{\infty}
\left [ n_2^2 + (n_1+\alpha)^2 t^2 ) \right ]^{-s}
\quad ,
\label{eq:nfpros}
\end{equation}
and identical statements can be made about its convergence properties
as in the previous subsection.
The $n_2$ summation in $S_{\rm ann}$ is carried out using a
contour integral representation
identical to that in Eq.\ (\ref{eq:contoureiget})
except that $x$$=$$(n_1+\alpha)t$.
The $n_1$ summation following the analog of Eq.\ (\ref{eq:contoureiget})
can be recognized as the Riemann zeta function
with two arguments,
\begin{equation}
 \sum_{n_1 = 0}^{\infty}
(n_1 + \alpha )^{-2s+1} t^{-2s+1} \equiv \zeta(2s-1,   \alpha) t^{-2s + 1}
\quad .
\label{eq:zetaa}
\end{equation}
Taking the $s$-derivative followed by the $s$$=$$0$ limit gives,
\begin{eqnarray}
\lim_{s\to 0} {{d}\over{ds}}&&
 \zeta(2s-1,  \alpha)t^{-2s+1} B(\half , s- \half )
\nonumber \\
=&&\lim_{s\to 0} {{d}\over{ds}}
 \zeta(2s-1,  \alpha) t^{-2s+1} {{{\rm sin} (\pi s)}\over{{\sqrt{\pi}}}}
\Gamma (1-s) \Gamma (s - \half )
= - 2 \pi t \zeta (-1,  \alpha)
\quad .
\label{eq:finttranseigka}
\end{eqnarray}
Substituting the relation $\zeta(-n,q)$$=$$-B_{n+2}^{\prime}(q)/(n+1)(n+2)$,
and combining the contributions for $q$$=$$\alpha $, and $1$$-$$\alpha$,
gives:
\begin{equation}
{{\pi t }\over{3}}  \left [ B_3^{\prime} (1- \alpha)
+ B_3^{\prime} ( \alpha)) \right ]
= 2 \pi t
\left [ \alpha^2 + {{1}\over{6}} - \alpha \right ]
\quad ,
\label{eq:tota}
\end{equation}
where we have used $B_n^{\prime}(q)$$=$$nB_{n-1}(q)$, and
$B_2(q)$$=$$q^2$$-$$q$$+$${{1}\over{6}}$.

\vskip 0.1in
Next, we tackle the non-constant pieces from the square brackets in
the analog of
Eq.\ (\ref{eq:contoureiget}). It is helpful to take the $s$-derivative and
the $s$ $\to$ $0$ limit prior to performing the contour integral.
We begin with the contour integral in the upper half-plane:
\begin{eqnarray}
I_2 ( x, s) =&&
\int_{{\cal C}_+} dz
 {{e^{ i\pi z}}\over{e^{i\pi z}-e^{-i \pi z}}}
  \left ( z^2 + x^2 \right )^{-s}
\nonumber \\
        =&&
- 2 {\rm sin} (\pi s) \int_{x }^{\infty} dy
 {{e^{ -\pi y  }}\over{e^{\pi y } -e^{- \pi y } }}
  \left ( y^2 - x^2 \right )^{-s}
         \quad .
\label{eq:twfsumtrans2eig}
\end{eqnarray}
Taking the derivative with respect to $s$, and setting $s$$=$$0$,
gives:
\begin{eqnarray}
{{d}\over{ds}} I_2 ( x, s)|_{s=0} =&&
- 2 \pi \int_{ x }^{\infty} dy
 {{e^{ -\pi y  }}\over{e^{\pi y } -e^{- \pi y } }}
\nonumber \\
=&& - {\rm log} \left ( 1 +e^{-2\pi x} \right )
\quad .
\label{eq:twequaeig}
\end{eqnarray}
The ${\cal C}_-$ integral gives an identical contribution as
before.
Combining the contributions to $S_{\rm ann}$
from terms with $\alpha$$=$$\phi/\pi$, and $1$$-$$\alpha$,
respectively,
gives the following result in the $m$$\to$$0$ limit:
\begin{equation}
S_{\rm ann} =  2 \pi t ( \alpha^2 + {{1 }\over{6}} - \alpha )
- 2 \sum_{n_1=0}^{\infty} {\rm log} \left [ (1+ e^{-2 \pi (n_1 + \alpha ) t })
 ( 1+ e^{-2 \pi (n_1 + 1 - \alpha ) t }) \right ]
\quad ,
\label{eq:equsum}
\end{equation}
The result for the functional determinant of the Laplacian acting
on a twisted complex scalar takes the form:
\begin{equation}
\left [
\prod_{\pm} \prod_{n_1=0}^{\infty} \prod_{n_2=-\infty}^{\infty \prime}
 \omega_{n_2 , n_1 } \right ]^{-1}
= q^{ \half ( \alpha^2 + {{1}\over{6}} -  \alpha) }
 (1-q^{\alpha})^{-1}
 \prod_{n_1=1}^{\infty}
\left [ ( 1- q^{n_1  - \alpha  })( 1- q^{n_1 + \alpha  } ) \right ]^{-1}
\quad ,
\label{eq:result}
\end{equation}
where $q$$=$$e^{-2\pi t}$. The result can be expressed in terms of the Jacobi
theta function as follows:
\begin{equation}
{{q^{\half \alpha^2 +{{1}\over{24}} - {{1}\over{8}}} }\over{ - 2i {\rm Sin} (\pi t \alpha/2)}}
 \prod_{n_1=1}^{\infty}
\left [ ( 1- q^{n_1  - \alpha  })( 1- q^{n_1 + \alpha  } ) \right ]^{-1}
 = - i {{ e^{\pi t \alpha^2} \eta(it)}\over{ \Theta_{11} (it \alpha ,it) }}
\quad ,
\label{eq:resulttat}
\end{equation}
with $\alpha$$=$$\phi/\pi$.

\vskip 0.1in
Setting $\alpha$$=$$0$ in this expression, and combining with the result in
Eq.\ (\ref{eq:zerota}), gives the functional determinant of the Laplacian
acting on a free Neumann scalar:
\begin{equation}
\left ( \prod_{n_2=-\infty}^{\infty \prime} \prod_{n_1 = 0}^{\infty}
\omega_{n_1 n_2} \right )^{-1/2} =
\left ({{1}\over{2t}} \right )
 q^{ -{{1}\over{24}}}
\prod_{n_1=1}^{\infty}
( 1- q^{n_1 } )^{-1}
= {{1}\over{2t}} \left [ \eta (it) \right ]^{-1}
\quad ,
\label{eq:resultd}
\end{equation}
where $q$$=$$e^{-2\pi t}$. The expression for the Dirichlet determninant
is identical except for the absence of the overall factor of $1/2t$.

\subsection{Mobius Strip: Twisted Complex Fermion Eigenspectrum}

\vskip 0.1in
As a final illustration, we work out the contribution from worldsheet
fermions to the Mobius strip amplitude for type I string theory in a
background magnetic field.
As explained in the text, we must identify the scalar eigenspace
inferred by application of Bose-Fermi equivalence. Under the
action of $\Omega$, the eigenspace in the $(a,b)$ fermionic sector
of the theory for the corresponding complex scalar takes the form:
\begin{eqnarray}
\Psi_{n_2  n_1 } =&& {{1}\over{{\sqrt{2t}}}}
 e^{4\pi i (n_2 + \half + \half (1\pm b)) \sigma^2}  {\rm Sin} 2\pi
(n_1 + \half + \half(1\pm a) + \half \alpha )\sigma^1
\nonumber\\
\Psi_{n_2  n_1 } =&& {{1}\over{{\sqrt{2t}}}}
 e^{4\pi i (n_2 +\half (1\pm b) )\sigma^2}  {\rm Sin} 2\pi
(n_1 + \half (1\pm a) + \half \alpha )\sigma^1
\quad ,
\label{eq:mssetf}
\end{eqnarray}
where we will set $\alpha$ equal to $\phi/\pi$, and $1$$-$$\phi/\pi$.
Note that eigenfunctions of odd (even) mass level are
weighted differently in the trace. The corresponding eigenvalues are:
\begin{eqnarray}
\omega^{\rm odd}_{n_2  n_1 } =&&
 {{4\pi^2}\over{t^2}} (n_2+\half + \half (1\pm b) )^2 + 4\pi^2 \left ( n_1+ \half
+ \half (1\pm a) + \half \alpha \right )^2
\nonumber\\
\omega^{\rm even}_{n_2  n_1 } =&&
 {{4\pi^2}\over{t^2}} (n_2 + \half (1\pm b))^2 + 4\pi^2 \left ( n_1 + \half (1\pm a)
  + \half \alpha \right )^2
\quad ,
\label{eq:mseisetf}
\end{eqnarray}
where $-\infty$$\le$$n_2$$\le$$\infty$, and $0$$\le$$n_1$$\le$$\infty$.
We will compute the product over both sets of
eigenvalues and then take the square root of the result. This gives:
\begin{eqnarray}
{\rm det} \Delta_{\rm mob }^{(a,b)} =&&
\prod_{n_2=-\infty}^{\infty} \prod_{n_1=0}^{\infty}
\left [ {{ 4\pi^2}\over{t^2}} (n_2 + \half (1\pm b))^2 + 4\pi^2 \left (n_1
+ \half (1\pm a)
+ \half \alpha \right )^2 \right ]^{1/2}
\nonumber\\
\quad &&\quad \times
\left [ {{ 4 \pi^2}\over{t^2}} (n_2 + \half + \half (1\pm b) ) ^2 +
  4\pi^2 \left (n_1 + \half + \half (1\pm a) + \half \alpha \right )^2
\right ]^{1/2}  .
\label{eq:mseigsnvf}
\end{eqnarray}
Denoting $\phi/\pi$ by $\alpha$, this gives the results:
\begin{eqnarray}
{\rm det} \Delta_{\rm mob}^{(0,0)} =&&
 = + i e^{-2 \pi t \alpha^2} [{{\eta(2it)}\over{ \Theta_{11} (2it \alpha ,2it) }}]^{-1/2}
  [{{\eta(2it)}\over{ \Theta_{00} (2it \alpha ,2it) }}]^{-1/2}
\nonumber\\
{\rm det} \Delta_{\rm mob}^{(0,1)} =&&
 = -  e^{-2 \pi t \alpha^2} [{{\eta(2it)}\over{ \Theta_{10} (2it \alpha ,2it) }}]^{-1/2}
  [{{\eta(2it)}\over{ \Theta_{01} (2it \alpha ,2it) }}]^{-1/2}
\nonumber\\
{\rm det} \Delta_{\rm mob}^{(1,0)} =&&
 = +  e^{-2 \pi t \alpha^2} [{{\eta(2it)}\over{ \Theta_{01} (2it \alpha ,2it) }}]^{-1/2}
  [{{\eta(2it)}\over{ \Theta_{10} (2it \alpha ,2it) }}]^{-1/2}
\nonumber\\
{\rm det} \Delta_{\rm mob}^{(1,1)} =&&
 = - i e^{-2 \pi t \alpha^2} [{{\eta(2it)}\over{ \Theta_{11} (2it \alpha ,2it) }}]^{-1/2}
  [{{\eta(2it)}\over{ \Theta_{00} (2it \alpha ,2it) }}]^{-1/2}
\quad .
\label{eq:msresultf}
\end{eqnarray}
Thus, allowing for the phase in front of each contribution, the contribution
to the Mobius strip amplitude from four complex worldsheet fermions, one of which is
twisted, takes the form:
\begin{eqnarray}
{\rm det} \Delta^{F}_{\rm mob}
 =&&   e^{-2 \pi t \alpha^2}
\left [{{\eta(2it)}\over{ \Theta_{01} (2it \alpha ,2it) }}
{{\eta(2it)}\over{ \Theta_{10} (2it \alpha ,2it) }} \right ]^{-1}
  \left ( {{\eta(2it)}\over{ \Theta_{01} (2it \alpha ,2it) }}
  {{\eta(2it)}\over{ \Theta_{10} (2it \alpha ,2it) }}\right )^{-3}
\nonumber\\
&& \quad  -
    e^{-2 \pi t \alpha^2}
\left [{{\eta(2it)}\over{ \Theta_{10} (2it \alpha ,2it) }}
{{\eta(2it)}\over{ \Theta_{01} (2it \alpha ,2it) }} \right ]^{-1}
  \left ( {{\eta(2it)}\over{ \Theta_{10} (2it \alpha ,2it) }}
  {{\eta(2it)}\over{ \Theta_{01} (2it \alpha ,2it) }}\right )^{-3}
,
\nonumber \\
\label{eq:msfermi}
\end{eqnarray}
where $\alpha$ denotes $\phi/\pi$.

\section{Off-shell Propagator of a Closed String}

\vskip 0.1in Let us return to our review in appendix A.2 of the
contribution to the one-loop amplitude of bosonic open and closed
string theory from world-surfaces with the topology of a cylinder.
From the perspective of the closed string channel, this graph
represents the tree-level propagation of a single closed string,
exchange between a spatially-separated pair of Dpbranes. A crucial
observation is as follows: although the Dpbrane vacuum corresponds
to a spontaneous breaking of translation invariance in the bulk
$25$$-$$p$ dimensional space orthogonal to the pair of Dpbranes,
notice that spacetime translational invariance is preserved within
the $p$$+$$1$-dimensional worldvolume of each Dpbrane.

\vskip 0.1in It is interesting to ask whether it is possible to
modify this calculation such that {\em all} $26$ spacetime
translation invariances are broken. We emphasize that we ask this
question not only for the point-like boundary limit of the annulus
graph, but for the annulus with {\em macroscopic} boundary loops.
The former limit with pointlike boundaries corresponds to the
tree-level exchange of a closed string between a pair of
Dinstantons: their worldvolumes are spacetime points, and each
boundary of the annulus is therefore mapped to a point in the
embedding $26$d spacetime. The latter case corresponds to a
genuinely new worldsheet amplitude, and the corresponding analysis
of the covariant string path integral brings in many new features,
first described in \cite{cmnp,wils}.

\vskip 0.1in It is convenient to align the macroscopic loops,
${\cal C}_i$, ${\cal C}_f$, which we will choose to have the
common length $L$, such that their distance of nearest separation,
$R$, is parallel to a spatial coordinate, call it $X^{25}$. As in
appendix A.2, the Polyakov action contributes a classical piece
corresponding to the saddle-point of the quantum path integral:
the saddle-point is determined by the minimum action worldsurface
spanning the given loops ${\cal C}_i$, ${\cal C}_f$. The result
for a generic classical solution of the Polyakov action was given
in \cite{cmnp}. For coaxial circular loops in a flat spacetime
geometry, we have a result identical to that which holds for a
spatially separated pair of generic Dpbranes in flat spacetime,
namely:
\begin{equation}
S_{\rm cl} [G,g] = \quaft \int d^2 \sigma {\sqrt{g}} g^{ab} ~
G^{25,25} (X)
   \partial_a X^{25} \partial_b X^{25}
 =  {{1}\over{2\pi \alpha^{\prime}}} R^2 t \quad .
\label{eq:backgd}
\end{equation}
Notice, in particular, that there is no $L$ dependence in the
saddle-point action as a consequence of the Dirichlet boundary
condition on all $26$ scalars. The metric on the annulus is
parameterized as before by a single real worldsheet modulus, $t$,
and it takes the form:
\begin{equation}
ds^2 = e^{\phi} ( (d\sigma^1)^2 + 4t^2 (d \sigma^2)^2 ), \quad
{\sqrt{g}} = e^{\phi} 2t , \quad 0 \le t \le \infty \quad ,
\label{eq:metricab}
\end{equation}
with worldsheet coordinates, $\sigma^a$, $a$$=$$1,2$,
parameterizing a square domain of unit length. $2t$ is the
physical length of either boundary of the annulus, as measured in
the two-dimensional field theory. As in the case of the
Dinstanton, we will evaluate the determinant of the scalar
Laplacian for all $26$ embedding coordinates with the Dirichlet
boundary condition. In addition, we should note that there is no
contribution from coordinate zero modes, since all of the
$X^{\mu}$ are Dirichlet. Thus, the usual box-regularized spacetime
volume dependence originating in the Neumann sector is absent,
precisely as in the vacuum of a pair of Dinstantons. The analog of
Eq.\ (\ref{eq:gaussxa}) reads:
\begin{equation}
1 = \int d\delta X  e^{-\half |\delta X|^2 } =
 \int d\delta X^{\prime} e^{-|\delta X^{\prime} |^2/2 }
 \quad .
\label{eq:gaussxam}
\end{equation}

\vskip 0.1in The crucial difference in the path integral
computation when the boundaries of the annulus are mapped to {\em
macroscopic} loops in embedding spacetime has to do with the
implementation of boundary reparametrization invariance: we must
include in the path integral a sum over all possible maps of the
worldsheet boundary to the loops ${\cal C}_i$, ${\cal C}_f$
\cite{cmnp}. Notice that the analysis of reparametrization
invariance in the bulk of the worldsheet is unaltered. As a
consequence, the conditions for Weyl invariance, and for the
crucial decoupling of the Liouville mode, are unchanged. The path
integral computation we are about to perform simply yields the
one-loop amplitude of the open and closed bosonic string theory
{\em in a distinct vacuum}. We will understand the nature of the
new boundary state characterizing this vacuum in a moment.

\vskip 0.1in Let us proceed with the analysis of the measure
following the steps in appendix A.2. The differential operator
mapping worldsheet vectors, $\delta \sigma^a$, to symmetric
traceless tensors, usually denoted $(P_1 \delta \sigma)_{ab}$, has
only one zero mode on the annulus. This is the constant
diffeomorphism in the direction tangential to the boundary:
$\delta \sigma^2_0$. The analysis of the diffeomorphism and Weyl
invariant measure for moduli follows precisely as for the annulus
\cite{mine}. The only difference is an additional contribution
from the vector Laplacian, accounting for diffeomorphisms of the
metric which are nontrivial on the boundary \cite{cmnp}. The
analog of Eq.\ (\ref{eq:jacoa}) now takes the form:
\begin{eqnarray}
1 =&& \int d e \int d g d X e^{-\half |\delta g|^2 - \half |\delta
X|^2 - \half |\delta e|^2} \cr =&& \left ( {\rm det}  Q_{22}
\right )^{-1/2}
  \int d^2 \sigma {\sqrt{g}}
\left ({\rm det}^{\prime} {\cal J} \right )^{1/2} \left ({\rm
det}^{\prime} {\cal M} \right )^{1/2}  \int
 (d\phi d\delta \sigma)^{\prime} d t  dX^{\prime}
           e^{-\half |\delta g|^2 - \half |\delta X|^2 - \half |\delta e|^2 }
           ,\cr
           &&
\label{eq:jacoab}
\end{eqnarray}
where $({\rm det} Q_{22})$ $=$ $2t$ in the critical dimension,
cancelling the factor of $2t$ arising from the normalization of
the integral over the single real modulus. As shown in
\cite{mine}, the functional determinant of the vector Laplacian
acting in the worldsheet bulk takes the form:
\begin{equation}
\left ( {\rm det}^{\prime} {\cal M} \right )^{1/2} = \left ({\rm
det}^{\prime} ~ 2 ~ \Delta^c_d \right )^{1/2} \left (
{{1}\over{2t}} \right ) = {{ 1}\over{2}} (2t)^{-1} {\rm
det}^{\prime} \Delta = \half (2t)^{-1}
\prod_{n_2=-\infty}^{\infty} \prod_{n_1=-\infty}^{\infty \prime }
\omega_{n_2 , n_1} \quad , \label{eq:vectab}
\end{equation}
and the infinite product is computed precisely as in appendix A.2.
The functional determinant of the operator ${\cal J}$ can likewise
be expressed in terms of the functional determinant of the
Laplacian acting on free scalars on the one-dimensional boundary,
parametrized here by $\sigma^2$ \cite{cmnp}. Thus, for boundary
length $2t$, we have $\Delta_b$$=$$(2t)^{-2}\partial_2^2 $, with
eigenspectrum:
\begin{equation}
\omega_{n_2 } = {{\pi^2}\over{t^2}} n_2^2  , \quad \Psi_{n_2} =
{{1}\over{{\sqrt{2t}}}}
 e^{2\pi i n_2 \sigma^2}
\quad , \label{eq:deegtermafb}
\end{equation}
where the subscripts take values in the range
$-\infty$$\le$$n_2$$\le$$\infty$.

\vskip 0.1in Thus, the connected sum over worldsurfaces with the
topology of an annulus with boundaries mapped onto spatially
separated macroscopic loops, ${\cal C}_i$, ${\cal C}_f$, of common
length $L$ takes the form \cite{cmnp,wils}:
\begin{equation}
 {\cal A}_{i,f} = \left [ L^{-1}(4\pi^2 \alpha^{\prime})^{1/2}
 \right ]
   \int_0^{\infty} {{dt}\over{2t}} \cdot (2t)^{1/2} \cdot
    \eta (it )^{-24} e^{-R^2 t /2\pi\alpha^{\prime}}
\quad . \label{eq:resulttafbs}
\end{equation}
The only change in the measure for moduli is the additional factor
of $(2t)^{1/2}$ contributed by the functional determinant of
${\cal J}$. The pre-factor in square brackets is of interest;
recall that there is no spacetime volume dependence in this
amplitude since we have broken translational invariance in all
$26$ directions of the embedding spacetime. If we were only
interested in the point-like off-shell closed string propagator,
as in \cite{cmnp}, the result as derived is correct without any
need for a pre-factor.\footnote{Comparing with the final
expression for the off-shell point-like propagator given in Eq.\
(4.5) of \cite{cmnp}, and letting $t$$\to$$2\lambda$ in order to
match with the notation in \cite{cmnp}, the reader should ignore
an extraneous factor of $\lambda^{-13}$, which should clearly be
absent in an all-Dirichlet string amplitude.} However, we have
{\em required} that the boundaries of the annulus are mapped to
loops in the embedding spacetime of an, a priori, fixed length
$L$. Since a translation of the boundaries in the direction of
spacetime parallel to the loops is equivalent to a boundary
diffeomorphism, we must divide by the (dimensionless) factor: $L
(4\pi^2 \alpha^{\prime})^{-1/2}$. This accounts for the pre-factor
present in our final result. Note that for more complicated loop
geometries, including the possibility of loops with corners, the
pre-factor in this expression will take a more complicated form.

\end{document}